\documentclass[aip,reprint,floatfix,onecolumn,footinbib]{revtex4-1}
\def\b#1{\mbox{\boldmath $#1$}}    
\usepackage{array,multirow,rotating}
\usepackage{graphics}
\usepackage{amsmath}
\usepackage{textcomp}
\usepackage{subfigure}
\usepackage{epsfig}

\newcommand{\dwat}{D$_2$O}
\usepackage{color}
\usepackage{lineno}

\begin{document}
\title{A Bayesian analysis of neutron spin echo data on polymer coated gold nanoparticles in aqueous solutions}
\author{Alessio De Francesco}
\altaffiliation{Consiglio Nazionale delle Ricerche, Istituto Officina dei Materiali c/o OGG Grenoble, France}
\author{Luisa Scaccia}
\altaffiliation{Dipartimento di Economia e Diritto, Universit\`a di Macerata, Via Crescimbeni 20, 62100 Macerata, Italy}
\author{R. Bruce Lennox}
\altaffiliation{Dep. Chemistry,
McGill University, Sherbrooke St. West, Montreal, Canada}
\author{Eleonora Guarini}
\altaffiliation{Dipartimento di Fisica e Astronomia, Universit\`a di Firenze, via G. Sansone 1, I-50019 Sesto Fiorentino, Italy}
\author{Ubaldo Bafile}
\altaffiliation{Consiglio Nazionale delle Ricerche, Istituto di Fisica Applicata  ''Nello Carrara'', via Madonna del Piano 10, I-50019 Sesto Fiorentino, Italy}
\author{Peter Falus}
\altaffiliation{Institut Laue-Langevin, Grenoble, France}

\author{Marco Maccarini}
\altaffiliation{Universit\'e Grenoble Alpes - Laboratoire TIMC/IMAG  UMR CNRS 5525 Grenoble, France}
\email{marco.maccarini@univ-grenoble-alpes.fr}

\begin{abstract}
We present a neutron spin echo study of the nanosecond dynamics of polyethylene glycol (PEG) functionalised nanosized gold particles dissolved in \dwat\ at two temperatures and two different PEG molecular weights (400D and 2000D). The analysis of the neutron spin echo data was performed by applying a Bayesian approach to the description  of time correlation function decays in terms of exponential terms, recently proved to be theoretically rigorous. This approach, which addresses in a direct way the fundamental issue of model choice in any dynamical analysis, provides here a guide to the most statistically supported way to follow the decay of the Intermediate Scattering Functions $I(Q, t)$ by basing on statistical grounds the choice of the number of terms required for the description of the nanosecond dynamics of the studied systems. Then, the presented analysis avoids from the start resorting to a pre-selected framework and can be considered as model free. By comparing the results of PEG coated nanoparticles with those obtained in PEG2000  solutions, we were able to disentangle the translational diffusion of the nanoparticles from the internal dynamics of the polymer grafted to them, and to show that the polymer corona relaxation follows a pure exponential decay in agreement with the behavior predicted by coarse grained molecular dynamics simulations and theoretical models. This methodology has one further advantage: in the presence of a complex dynamical scenario $I(Q,t)$ is often described in terms of the Kohlrausch-Williams-Watts function that can implicitly represent a distribution of relaxation times. By choosing to describe the $I(Q,t)$ as a sum of exponential functions and with the support of the Bayesian approach, we can explicitly determine when a finer-structure analysis of the dynamical complexity of the system exists according to the available data without the risk of overparametrisation. The approach presented here is as an effective tool that can be used in general to provide an unbiased interpretation of neutron spin echo data or whenever spectroscopy techniques yield time relaxation data curves.
\end{abstract}
\maketitle

\section{Introduction}
Considerable scientific effort has been devoted in the last years to the study of gold nanoparticles (NP), in view of many important properties peculiar of these systems including quantum size effects and single electron transitions.\cite{Alivisatos1996}
Due to their small size (1-20 nm) the electronic bands of the nanoparticles follow  quantum-mechanical rules and exhibit a strong dependence of the physical properties on the size and shape of the particles.\cite{brust2002} These properties are further highly dependent upon the presence of an organic ligand coating or capping layer  and upon the spatial relationship of one particle to another. 

In order to contrast and control their tendency to aggregate, different methods have been developed to synthesise and stabilise such particles. Notable among these is the method developed by Brust {\it et al.}\cite{Brust1995} This method has had considerable impact since it has allowed the production of gold NP (AuNP) of controlled size and reduced size dispersity stabilised by alkanethiols. This has enabled  a series of functionalisations that allow using the  AuNPs in a number of applications including molecular recognition\cite{labande2000,lin2002} and specific binding to biomolecules\cite{reynolds2000} with significant impact in biological and biomedical studies.\cite{astruc2004,Dreaden2012}

Functionalisation  with thiol-terminated polymers provides great flexibility in modulating  the properties of the AuNPs. It was shown that AuNPs  protected with 4-(N,N-dimethylamino)pyridine (DMAP)  are a convenient precursor of thiol-capped AuNPs. DMAP can be conveniently exchanged with thiol-terminated polymers to obtain AuNP that are soluble in organic solvents or water by choosing the appropriate thiol-terminated polymer including poly(ethylene oxide) (PEG) or polystyrene (PS) over a range of molecular weights.\cite{rucareanu2008}

The properties of the organic capping layer that protects the AuNPs impact the way the AuNPs interact with other systems such as a surrounding polymer matrix,\cite{genix2015} other NPs in complex 3D architectures,\cite{Zhihong2010} or a biomimetic assembly.\cite{tatur2013} For practical applications it is crucial to have a very precise knowledge of the molecular details and properties of this protective layer surrounding the NP.

In a previous study, a structural characterisation of 5 nm PEG-functionalised AuNPs was achieved via a combination of experimental methods including neutron and X-ray small angle scattering, transmission electron microscopy, density, and thermogravimetric measurements.\cite{maccarini2010} The functionalisation was obtained through thiol derivatisation of PEG2000 (45 units). The confinement of the PEG molecule on the highly curved surface of a AuNP was found to considerably affect the conformation and the hydration of the polymer as a function of the distance from the surface of the gold core.

Here, we extend this static study to the dynamical properties of these PEG-functionalised AuNPs. The  study of the dynamical processes  in nanosized objects like micelles, proteins, and nanoparticle dispersions presents some difficulties due to the presence of multi-scale dynamics occurring at similar time and range of length scales. Methods like Dynamic light scattering (DLS) are useful to measure the translational diffusion of the whole object under study. However, since the wavelength of light is larger than the dimension of these nanosized objects, internal dynamics cannot be explored with this technique. Quasi elastic neutron scattering (QENS) and, in particular, neutron spin echo spectroscopy (NSE)  have been shown to be an excellent tool to probe the dynamical processes in micellar dispersions.\cite{matsuoka2000,kanaya2005,castelletto2003} Thanks to its wide dynamical range, NSE can probe a  variety of dynamical processes, from translational diffusion of the whole micelle to  the internal dynamics of the polymer chains. 

The polymer coated NPs used in this study have sizes of the same order of those of the micellar aggregates studied in other investigations (50-200 \AA).\cite{matsuoka2000,kanaya2005,castelletto2003} However, they differ significantly in various aspects. Firstly, unlike the core of a micellar aggregate, the gold core of the NP is a solid. The internal dynamics of gold in the solid state include atomic vibrations and phonons that occur at much faster timescales than those detectable with NSE. Hence, differently from the micellar aggregate comparators, in the systems reported here the  dynamical processes present in the polymer corona and in the NPs core occur at significantly different timescales. Moreover, the nonionic surfactants, or block co-polymers of the micelles, aggregate in response to the hydrophobic effect, which favours the confinement of the hydrophobic moieties in the oily bulk. These aggregates are dynamic entities, as the individual surfactant molecules have a finite solubility in the solvent phase and can leave an aggregate to become part of another one, following a kinetic equilibrium.  PEG-grafted gold NPs, on the other hand, involve a strong Au-S bond (ca. 160-200 kJ/mol) between the thiol-derivatised  PEG to the metal surface. The exchange of material between different NPs, if it in fact occurs, does so at an exceedingly lower rate under ambient condition. NSE was also used to study NPs with various functionalisations dispersed in polymeric materials to elucidate the effect of the NP dispersion on the dynamical properties of the NPs-polymer composite.\cite{senses2016, senses2018,campanella2016,nusser2011,mark2017} These studies differ from that reported here in that the NPs are dispersed in a high molecular weight polymer matrix, whereas the NPs used in our experiments were dispersed in a low molecular weight solvent.

The analysis of the NSE data was undertaken by adopting an approach composed by two building blocks:
\begin{enumerate}
\item we expressed the NSE intermediate scattering function $I(Q,t)$ as a sum of exponentials following the results of general theoretical work  that proves that any time correlation function can be expressed as the sum of infinite (generally complex) exponential functions\cite{Barocchi_2012, Barocchi_2013, Barocchi_2014}. This provided an unbiased model-free expression to interpret the experimental NSE results.
\item The analysis of the $I(Q,t)$  was supported by a Bayesian approach implemented through the use of a Monte Carlo Markov Chain (MCMC) algorithm integrated with a Reversible Jump (RJ)\cite{Green95} option and successfully applied recently to Brillouin inelastic neutron scattering \cite{DeFrancesco2016}  and  inelastic X-ray scattering\cite{DeFrancesco2018} data.
\end{enumerate}

This methodology addresses one of the most important issues in the analysis of inelastic and quasi-elastic neutron scattering data - the arbitrariness in the choice of the number of components needed  to describe  the experimental dynamic structure factor. In NSE this issue translates in how many components one chooses to describe the decay of the intermediate scattering function (the time-Fourier transform of the dynamic structure factor). In this  approach the number of components is treated as a parameter to be estimated. Since inferential results from this approach are in the form of probabilities, the final choice of the number of components resides on a solid probabilistic basis. The main objective of this work is then to show that the combination of points 1 and 2 above allows a precise and theory complying representation of macromolecular polymeric systems in solutions and yields an unbiased and model-free description of the dynamics of these systems.

Furthermore, it is common practice in the analysis of NSE data to use a stretched exponential function to model the decay of the intermediate scattering function. The use of this function is  in some circumstances a heuristic way to account for an ensemble of dynamical processes occurring in the sample with a given distribution of relaxation times and the stretching parameter might measures, in some fashion, the difficulty in describing such a complex dynamics without however being able to give account for this complexity. It is also the results  of model based theoretical description of polymer dynamics such as the Zimm or the Rouse Model. Nevertheless,  all possible descriptions of the  experimental $I(Q,t)$ of the system under study should eventually satisfy the above general property.\cite{Barocchi_2012, Barocchi_2013, Barocchi_2014} This principle has been exploited to give excellent descriptions of time correlation functions and spectra relevant to simpler liquids.\cite{bellissima,bellissima2,guarini_au}  Here, by combining it with the  Bayesian approach, we  analyse the data in an alternative unbiased and model-free way by assigning  a  {\it fine-structure } of relaxation processes to an otherwise undistinguished representation of the decay of the time correlation function as it is provided by the beta-stretching modelling.  Furthermore and most importantly, the Bayesian method  used here is able to provide the most statistically reliable description of the system supported by the experimental data without any risk of over-parametrisation. 

We will also show that our unbiased method will provide results for the polymer corona in agreement with those predicted by coarse grained MD simulations and traditional theoretical models of polymer science.

\section{Materials and methods}

Polyethylene glycol coated gold nanoparticles (PEG AuNPs) were synthesised with the same procedure described in our previous work. \cite{maccarini2010} Briefly, tetraoctylammonium bromide-Au and DMAP-AuNPs were synthesised using Gittins and Caruso's procedure. \cite{gittins:2001,gandubert:2005} The Au-NPs were subjected to place-exchange reaction of presynthesised DMAP stabilised NPs with thiol-functionalised PEG2000SH and PEG400SH. The gold core had a radius around 25 \AA\ with a polydispersion of 20\% as a result of TEM analysis.\cite{rucareanu2008} However the indetermination in the Au core radius as a result of  small angle neutron scattering was around 2\%.\cite{maccarini2010} The detailed synthesis and characterisation are described elsewhere. \cite{rucareanu2008} PEG2000SH was purchased from Polymer Source Inc. Dorval, QC, Canada (M$_w$ $\sim$ 2000) and PEG400SH (M$_{w}$ $\sim$ 400) was received from Polypure Oslo, Norway. 
PEG2000 and PEG400 homopolymers were purchased from Sigma and used without further purification. The translational diffusion coefficients of the NPs were obtained by DLS. These measurements were carried out using an ALV CGS-3 Compact Goniometer equipped with a HeNe Laser with a wavelength of 632.8 nm, a 22 mW output power, and an ALV LSE-5004 Correlator. Samples were measured at a scattering angle of 90$^{\circ}$, while the sample temperature was controlled via an external water bath circulator. The small angle neutron scattering profile for the PEG2000 AuNP was measured on the D11 SANS instrument at the Institut Laue-Langevin with a standard configuration. The neutron wavelength was 6 \AA\ and two detector distances (10.5 and 1.5 m) were used. This provided a \textit{Q} range between $6 \cdot 10^{-3}$ and 0.43 \AA$^{-1}$ The PEG2000 AuNPs were dissolved in \dwat\ at a concentration of 1\% in weight.


\begin{figure}[h]
	\centering
	\includegraphics[width=0.8\linewidth]{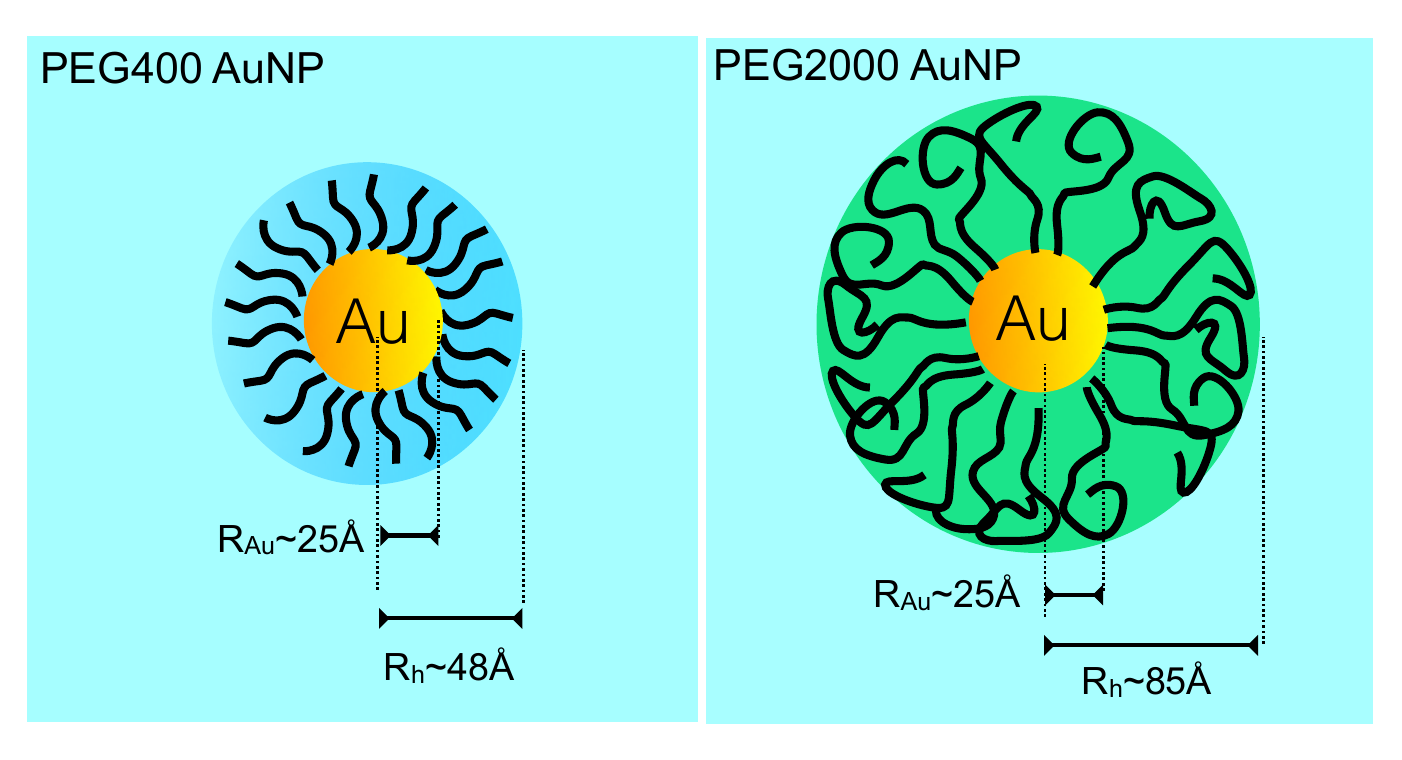}
	\caption{Sketch of the PEG400 and PEG2000 functionalised NPs. An indication of the characteristic sizes of the Au core (R$_{Au}$) and of the hydrodynamic radius $R_h$ are also shown.}
	\label{fig:aunp}
\end{figure}

\subsection{Neutron Spin Echo Spectroscopy}

NSE experiments were performed at the IN15 and IN11 neutron spin echo spectrometers at the Institut Laue-Langevin in Grenoble. NSE measurements were carried out in solutions of PEG2000-SH and PEG400-SH coated nanoparticles dispersed in D$_2$O at concentrations of 5\% wt and 2\% wt, respectively, and in solutions of PEG2000 homopolymers (10\% wt) and of PEG400 homopolymers (5\% wt), also in  D$_2$O. The NSE signal of  D$_2$O alone was also measured in the same experimental conditions to allow for background subtraction. All measurements of $I(Q,t)$  were performed in a wide range of $Q$ and time at two temperatures, 280 and 318 K. Different setups were used depending on the sample and the spectrometer. A complete summary of the measured samples and experimental setups is given in  Table \ref{tab:setup}.

 \begin{table}[h]
\centering
	\caption{Instrumental setups used for the various samples}
		\begin{tabular}{|c|c|c|c|c|c|c|} \hline
		 Sample       & $C$[w\%] &Instrument&$\lambda$ [\AA]&NSE time [ns]&$Q$ range [\AA$^{-1}$]\\ \hline
		 PEG2000 AuNP & 5  &   IN15  & 10, 16    &    0.3-206      &   0.02-0.21  \\ \hline
		 PEG400 AuNP  & 2  &   IN15  & 8,  16    &    0.18-206     &   0.02-0.25  \\ \hline
		 PEG2000      & 10 &   IN15  & 8,  16.8  &    0.18-240     &   0.02-0.25  \\ \hline
		 PEG2000      & 10 &   IN11C & 8.5       &    0.02-5       &   0.09-0.27  \\ \hline
		 PEG400       & 5  &   IN15  & 8,  16    &    0.18-206     &   0.02-0.25  \\ \hline

		\end{tabular}\label{tab:setup}
\end{table}

\subsection{Theory }\label{sec:theory}
 We briefly present in this section the expressions on which are based the models adopted for analysing the neutron intermediate scattering functions (nISF) $I(Q,t)$ measured in the neutron spin echo experiments. 
 
A number of customary approximation  is generally adopted to obtain a workable  expression of $I(Q,t)$. First of all the concentration of the nanoparticles in the solutions is supposed to be low enough to neglect any correlation between atoms of different nanoparticles. Then it is assumed that translational, rotational and vibrational motions are decoupled. This hypothesis is strictly valid for very dilute solutions but it is frequently used for systems at higher concentrations\cite{hui2017} and quite plausible for our system as we will show later. 
Under these hypotheses it can be shown that the nISF, after subtraction of the measured solvent contribution, can be expressed as:

\begin{equation}\label{Ifurthersimple}
{I}(Q,t)\simeq \exp(-DQ^{2}t) \frac{1}{N_{at}}\sum_{n,n'=1}^{N_{at}}\bigl\langle b_{n}^*b_{n'}\bigr\rangle\langle \exp[i\vec{Q}\cdotp(\vec{r}_{n}(t)-\vec{r}_{n'}(0))]\rangle= \exp(-DQ^{2}t) {I}_{\rm int}(Q,t) 
\end{equation}

where $D$ is the translational diffusion coefficient of the macromolecule, $N_{at}$ is the total number of atoms in a nanoparticle, $b_n$ is the scattering length of the $n$-th atom, and $r_n(t)$ is the displacement of an atom from its equilibrium position due to vibration or localised diffusive motion\cite{Kneller2004}, so that the last factor, labeled with the subscript `int', describes the internal dynamics of the nanoparticle.  As far as rotation are concerned it is useful to recall that the combination of independent translational and rotational motions has been shown\cite{Sharma2012} to be satisfactorily described in the frequency domain by a single central Lorentzian line, though with a slightly increased width with respect to Fick's Lorentzian of pure translational diffusion. This suggests that rotations can be accounted for by an effective, small, increase of the translational diffusion coefficient. However, in the specific case of the present samples, as reported in section \ref{sec:experimental}, no such increase has been detected, suggesting that the rotational contribution to Eq. (\ref{Ifurthersimple}) is negligible. 

 
In liquid samples the position of a nucleus as $t \to \infty$ becomes uncorrelated with its initial value. Conversely, if the motion of the nucleus is well located in space, or confined to a particular volume (e.g. reorientations of chemical groups in a molecule), the correlation persists and tends to a nonvanishing asymptotic value at long times. The latter case applies to the internal dynamics of the polymer molecules grafted to the nanoparticles, therefore also in the present case it is possible, at any time, to write ${I}_{\rm int}(Q,t)$ as the sum of its time-independent and time-dependent parts:

\begin{equation}
\label{boh}
{I}_{\rm int}(Q,t)={I}_{\rm int}(Q,\infty)+{I}^{'}_{\rm int}(Q,t).
\end{equation}

\noindent Here, the term ${I}_{\rm int}(Q,\infty)$ represents a quantity proportional to the contribution of  purely elastic scattering to the differential scattering cross section\cite{Squire} and is related to the structure of the nanoparticle. 

 NSE experiments  measure  a normalized intermediate scattering function (nISF),  which is normally written as:

\begin{equation}\label{ISFNSE}
\frac{{I}(Q,t)}{{I}(Q,0)}=\frac{{I}^{coh}(Q,t)-\frac{1}{3}{I}^{incoh}(Q,t)}{{I}^{coh}(Q,0)-\frac{1}{3}{I}^{incoh}(Q,0)}
\end{equation}

\noindent where  in general:

\begin{equation}\label{NSEcoerente}
{I}^{coh}(Q,t)=\frac{1}{N_{at}}\sum_{\substack{n,n'=1}}^{N_{at}} \langle b_n^{*coh}b_{n'}^{coh}\rangle\langle e^{-i\vec{Q}\cdot\vec{r_n}(0)}e^{+i\vec{Q}\cdot\vec{r_{n'}}(t)}\rangle
\end{equation}

\begin{equation}\label{NSEincoerente}
{I}^{incoh}(Q,t)=\frac{1}{N_{at}}\sum_{n=n'=1}^{N_{at}}\langle (b_n^{incoh})^2\rangle\langle e^{-i\vec{Q}\cdot\vec{r_n}(0)}e^{+i\vec{Q}\cdot\vec{r_n}(t)}\rangle
\end{equation}

Both these terms are implicitly contained in  Eq. (\ref{Ifurthersimple}) and contribute to $I(Q,t)$. In the experiment performed here  the coherent term prevails especially at low \textit{Q} ($\le$ 0.1 \AA$^{-1}$). In the region of higher \textit{Q}  the incoherent terms reached 40-50\% of the total contribution as it is shown in Figure S 1.

To further model ${I}^{'}_{\rm int}(Q,t)$, we refer to the already quoted general theoretical results \cite{Barocchi_2012, Barocchi_2013, Barocchi_2014}, showing that any correlation function that decays to zero at long times is exactly represented by an infinite series of exponentials of generally complex argument, each exponential being weighted, in the series, by a generally complex amplitude. When the amplitude and time constant of an exponential term are complex, then the sum of such term with its complex conjugate (also present in the series) describes an exponentially damped oscillation. Differently, if amplitude and time constant are real, then the exponential term depicts a pure exponential decay.\footnote{The finding that discrete dynamical modes of exponential type can be identified in an ISF is also present in the work of Kikuchi {\it et al.}\cite{kikuchi2013} However, this appears as the result of a numerical analysis rather than a property demonstrated in full generality.} It is important to note that the possibility of expressing the ${I}^{'}_{\rm int}(Q,t)$ as a sum of exponential terms is valid irrespectively of the fact that it represents, or is obtained from a coherent, incoherent scattering or a combination of the two.

Since we are dealing with quasielastic spectra, we can rather confidently assume that the normalized time correlation function ${I}^{'}_{\rm int}(Q,t)$ does not require complex modes in the series representation, and that it can be reliably described by real modes only. Moreover, in Refs \cite{bellissima,bellissima2,guarini_au}  it was shown that time correlation functions are most typically described with high accuracy by using only a small number (let's say $k$) of terms in the series, thus finally we modeled the internal dynamics as

\begin{equation}
\label{int}
\frac{{I}_{\rm int}(Q,t)}{{I}_{\rm int}(Q,0)}=\frac{{I}_{\rm int}(Q,\infty)}{{I}_{\rm int}(Q,0)}+(1-\frac{{I}_{\rm int}(Q,\infty)}{{I}_{\rm int}(Q,0)})\sum_{j=2}^{k} I_j \exp \Bigg(-\frac{t}{\tau_{\rm int, \it j}}\Bigg).
\end{equation} 

\noindent The reason why the index $j$ starts at 2 will be clear immediately below. In Eq. (\ref{int}) $\tau_{int,j}$ and $I_{j}$ are, respectively, the time constant and the amplitude of the $j$-th decay channel considered in the series (we omitted their obvious dependence on $Q$), and $\sum_{j=2}^{k} I_j =1$. The factor in front of the exponential series is imposed by the normalization condition $[{I}_{\rm int}(Q,t)/{I}_{\rm int}(Q,0)]_{t=0}=1$. Thus, by substituting Eq. (\ref{int}) in Eq.(\ref{Ifurthersimple}), our system can actually be modeled as 

\begin{equation}\label{IH_series} 
\frac{{I}(Q,t)}{{I}(Q,0)}\simeq \exp(-DQ^{2}t) \Bigg[ A_1+\sum_{j=2}^{k} A_j \exp \Bigg(-\frac{t}{\tau_{\rm int,\it j}} \Bigg) \Bigg],
\end{equation}

\noindent with $A_1=I_{\rm int}(Q,\infty)/I_{\rm int}(Q,0)$, $A_{j}=(1-A_1)I_{j}$ for $j\geq2$, and the sum rule $\sum_{j=1}^{k} A_j=1$. Defining $1/\tau_{1}=DQ^{2}$  and $1/\tau_{j}=1/\tau_{1}+1/\tau_{int,j}$ for $j\geq2$, Eq. (\ref{IH_series}) can  be re-written as

\begin{equation}
\label{IH_series2} 
\frac{{I}(Q,t)}{{I}(Q,0)}\simeq A_1\exp(-\frac{t}{\tau_{1}}) +\sum_{j=2}^{k} A_j \exp(-\frac{t 	}{\tau_{j}}).
\end{equation}

\noindent As we will see below in the results section, for the AuNP at the lowest values of $Q$ in the explored range, the decay of the time correlation function is fully explained by the translational  diffusive term ($A_{1}=1$) alone, therefore by what we will call from now on the relaxation time $\tau_{D}=\tau_{1}$. In the opposite limit of higher \textit{Q} (i.e. $QR_{h}>>1$, being $R_{h}$ the hydrodynamic radius) the weight $A_{1}$ progressively vanishes and the decay is more and more explained by the internal polymer relaxation dynamics. In the intermediate \textit{Q}-range the relaxation times $\tau_{j}$ in the second term of Eq. (\ref{IH_series2}) are somehow affected by the translational diffusive contribution of the particle.

 Since, as mentioned above,  the term $I_{\rm int}(Q,\infty)$  is related to the structure of the nanoparticle, the  term $A_1$ accounts implicitly for the form factor of the PEG AuNP.

In applying our RJ-MCMC approach (see next section), we will use for the AuNP the equation model (\ref{IH_series2}) where the relaxation times $\tau_{j}$ for $j\geq2$ will be explicitly referred to as $\tau_{pol,j}$. 
For the case of the homopolymer solutions of PEG2000 and PEG400, we used a more general expression of the nISF that reflects the fact that we did not make any assumption on the possible character of the first relaxation and that was instead inferred only after the data analysis:

\begin{equation}\label{eq:Intscatthomopol}
\frac{{I}(Q,t)}{{I}(Q,0)}\simeq \sum_{j=1}^{k} A_{j}\exp(-\frac{t}{\tau_{j}}).
\end{equation}

 It is worth noticing that the formulation of a model describing the relaxation dynamics in terms of simple exponential functions, as in Eqs. (\ref{IH_series2}) and (\ref{eq:Intscatthomopol}), is not the standard choice in polymer physics, where stretched exponentials are often assumed. In order to demonstrate how the intermediate scattering function for the considered systems can be fully described as a linear combination of simple exponentials, we generalised the model equations (\ref{IH_series2}) and (\ref{eq:Intscatthomopol}) to sums of Kohlrausch-Williams-Watts exponential functions with stretched coefficients $\beta_{j}$.  By  means of the Bayesian approach described in Section \ref{sec:bayes}  applied to this model, we will provide a quantitative estimate of the statistical significance of stretching parameters and we will show that the choice of non-stretched ($\beta_{j}=1$) exponentials is fully adequate to an accurate description of the studied system dynamics and indeed the most valid on a probabilistic basis.

\subsection{Bayesian analysis} \label{sec:bayes}

Equation (\ref{eq:Intscatthomopol}) describes  the measured nISF relaxation  as a linear combination of pure exponential components. As explained at the end of Section  \ref{sec:theory}, we generalised this model by replacing the simple with stretched exponential functions. 
Let $t=(t_1,\ldots,t_n)$ denote the vector of discrete times at which data were collected, and $y=(y_1,\ldots,y_n)$ the corresponding vector of the nISF measurements. Generalising Eq. (\ref{eq:Intscatthomopol}), the data can be assumed to be sampled from the following expression:
\begin{equation}\label{eq:datamodel}
y_i=\gamma\sum_{j=1}^{k} A_{j}\exp\left(-\left(\frac{t_i}{\tau_{j}}\right)^{\beta_j}\right)+\epsilon_i, \qquad \mbox{for } i=1,\ldots,n
\end{equation}
where $\gamma$ is a proportionality constant to account for a possible adjustement of the data normalisation and $\epsilon_i$, for  $i=1,\ldots,n$, are independent random noises, accounting for statistical errors in the measurements. We assume these random errors to be independent and identically distributed according to a normal distribution $\mathcal{N}(0,\nu\sigma^2_i)$, where $\sigma_i$ is the measurement error corresponding to the $i$-th observation and $\nu$ is a proportionality constant.

In order to choose an adequate number of components and estimate their parameters, we adopted a Bayesian approach. The main advantage of this approach in the present context is that the number of components, i.e. the relaxations in the time correlation function decay, can be treated as an unknown parameter to be estimated, along with the other model parameters. In addition, inferential results are obtained in the form of parameter probability density functions, so that  the final choice on the number of relaxations is justified on a probabilistic basis. Finally, the choice of the Bayesian approach intrinsically and automatically applies the Ockam's razor principle\cite{DeFrancesco2016}, thus preventing over-parametrisation.

Bayesian inference is based on Bayes theorem \citep{Bayes53}, which allows to combine evidence from data and possible prior knowledge $I$,  through the equation
\begin{equation}\label{BayesTeo}P(\Theta|y,I)\propto P(y|\Theta,I)\times P(\Theta|I),\end{equation}
where the symbol $|$ stands for ``conditional on'', and $\Theta$ is the whole vector of unknown parameters, i.e. $\Theta=(k, A,\tau,\beta,\gamma,\nu)$, with $A = (A_1,\ldots,A_k)$, $\tau = (\tau_1,\ldots,\tau_k)$ and $\beta = (\beta_1,\ldots,\beta_k)$.  We stress, again, that  the number $k$ of elementary components in the model is also treated as a parameter.

The first term on the right-hand side of Eq. (\ref{BayesTeo}) is the  likelihood function, which is the probability of the observed data $y$, conditional on a certain set $\Theta$ of parameter values, and on prior knowledge. In the present case the likelihood of the data is:
\begin{equation}\label{likelihood}
P(y|\Theta)=\prod_{i=1}^n\phi\left(y_i;f(t_i),\nu\sigma_i^2\right),\end{equation}
where $\displaystyle f(t_i)=\gamma\sum_{j=1}^{k} A_{j}\exp\left(-\left(\frac{t_i}{\tau_{j}}\right)^{\beta_j}\right)$ and $\phi\left(\b\cdot;f(t_i),\nu\sigma_i^2\right)$ is the density
of the $\mathcal{N}\left(f(t_i),\nu\sigma_i^2\right)$ distribution, evaluated in a specific point ``$\b\cdot$''. The conditioning on $I$ here and in the rest of the paper has been dropped to simplify the notation.

The second term on the right-hand side of Eq. (\ref{BayesTeo}) is the prior distribution for the parameters of the model function, before data collection. It incorporates all prior knowledge, if any, on the problem at hand. In a Bayesian perspective, such prior distributions need to be specified for each parameter of the model, including the unknown number of mixture components, $k$. 

In order to allow for complete  {\it a priori} ignorance on $k$, we assume $p(k)=1/k_{\max}$, for $k=1,\ldots,k_{\max}$, where $k_{\max}$ is reasonably chosen to be the maximum possible number of components.
The vector of weights, $A$, is assumed to follow {\it a priori} a Dirichlet distribution with hyperparameters $\lambda=(\lambda_1,\ldots,\lambda_k)$, i.e $A\sim\mathcal{D}(\lambda)$. Complete ignorance can be accommodated by letting $\lambda_1=\ldots=\lambda_k=1$ so that each component of the mixture has a marginal weight which is uniformly distributed in the interval $(0;1)$. Different prior beliefs on $k$ and $A$ can be easily accounted for\cite{DeFrancesco2016}.
The stretching parameters $\beta_j$ are assumed  {\it a priori} to independently follow a mixed distribution made up of a continuous beta density and a probability mass in 1, that is 
\begin{equation*}\beta_j\sim \zeta\mathcal{B}(\kappa,\psi)+(1-\zeta)\delta_{\beta_j,1},\end{equation*}
with $\kappa$ and $\psi$ being the hyperparameters of the beta density, and $\zeta$ denoting our prior support in favour of stretched (rather than unstretched) components. Notice that, letting $\zeta=0$ implies assuming a sum of pure exponential components, as the one in Eq. (\ref{eq:Intscatthomopol}), whereas letting $\zeta=1$ determines a sum of stretched components. For any other $0<\zeta<1$, the mixture can include both stretched and unstretched components.  In particular, we assumed $\zeta=0.5$, so that the probability of having a stretched or unstretched $j-$th component is, a priori, the same, for all $j=1,\ldots,k$. In addition, we set $\kappa=1$ and $\psi=1$ so that the stretching parameters are assumed uniformly distributed on the interval $(0;1)$. This corresponds to an uninformative prior giving the probability of 0.5 to an unstretched component and the remaining probability of 0.5 to a stretched component,  and, for a stretched component, assigning the same density to any value of $\beta_j$ in $(0;1)$. Therefore our choice is that of being completely general: the same prior probability is given to any model made up of a sum of just stretched components or a sum of just unstretched components or a sum of stretched and unstretched components. Obviously models with just one stretched or unstretched component are also included when $k=1$. Our choice of prior satisfies both the requirements of the theory of correlation functions \cite{Barocchi_2012, Barocchi_2013, Barocchi_2014} and the possibility to describe the results in terms of theoretical models  used in polymer physics\cite{richter2005} without pushing the data in either one or the other direction. Obviously, different priors can be considered. For example, if one wants to give a higher prior to the Zimm model, he could choose $\kappa$ and $\psi$ such that $\kappa/(\kappa+\psi)\approx 0.85$ (so that the a priori expected value of $\beta_j$ would be 0.85), with $\psi$ accounting for the degree of prior belief in the Zimm model: the smaller $\psi$, the higher the prior probability assigned to this model. If, instead, one wants to consider both Zimm and Rouse model as a priori more probable than other solutions, a mixture of two beta densities could be considered for $\beta_j$, with two peaks centered on 0.85 and 0.5, respectively, and mixture weights proportional to the prior probabilities assigned to the two models.  

 The relaxation times $\tau_j$, for $j=1,\ldots,k$ are assumed to be {\it a priori} independent and identically distributed according to uniform densities in the interval $[0;\tau_{\max}]$, with $\tau_{\max}$ reasonably chosen. 
For the parameter $\gamma$, we also assume a mixed distribution made up of a continuous beta density and a probability mass in 1, that is $$\gamma\sim \xi\mathcal{B}(\varphi,\rho)+(1-\xi)\delta_{\gamma,1},$$
with $\varphi$ and $\rho$ being the hyperparameters of the beta density, and $\xi$ denoting our prior support in favour of a refinement of the data normalisation process. As for $\zeta$, also in this case we let $\xi=0.5$. Finally,  for conjugacy reasons, $\nu$ is assumed to have an inverse Gamma prior,
$\nu^{-1}\sim \mathcal{G}(\iota,\varsigma)$, parametrised so that the mean and the variance are $\iota/\varsigma$ and $\iota/\varsigma^2$.

In the case of PEG AuNP data, where the first component of the model is meant to represent explicitly the centre-of-mass diffusion described as in Eq. (\ref{IH_series}), we specify slightly different priors. In fact, the stretching parameter is fixed at the value 1 by using a degenerate prior distribution $P(\beta_1 = 1) =1)$. Moreover, for the translational diffusion parameter, we assumed $D\sim \mathcal{N}(\mu_D,\sigma^2_D)$, with the hyperparameters chosen in such a way that $\mu_D$ is equal to the value $D_{LS}$ obtained by independent DLS   measurements (see Section \ref{sec:Peg}) and $\sigma_D = 0.20\mu_D$, so that $D$ can vary by up to $40\%$ from $D_{LS}$, with a probability of $95\%$. This was done to account for possible concentration effects on the translational diffusion, since the DLS experiment was done in a very dilute regime, whereas the NSE was done at a concentration of 5\% in weight. Notice that, the relaxation time of the first exponential component, conditional on $D$, is not random anymore and is simply obtained as $\displaystyle\tau_D=\frac{1}{DQ^2}$.

Bayesian inference is based on the joint posterior distribution of all the random parameters, i.e. the left-hand side in Eq. (\ref{BayesTeo}). Due to the complexity of the model at hand, such posterior distribution is known up to a normalisation constant, which would require solving a high-dimensional integral to be evaluated\cite{Andrieu01,Razul03} (i.e. the integral of the right-hand side of Eq. (\ref{BayesTeo}) over the parameters $\Theta$, which is required in the denominator of the right-hand side to have the joint posterior density integrate to one and the proportionality symbol replaced by equality). For this reason, we resort to MCMC algorithms to simulate the joint posterior distribution of interest.

MCMC algorithms allow to draw samples from a target distribution (the joint posterior in Bayesian applications), known up to a normalising constant. In this regard, these algorithms make it feasible to simulate the joint posterior distribution for the parameters of  arbitrarily complicated models. MCMC algorithms are based on the construction of an irreducible and aperiodic Markov chain that has the target distribution as invariant  distribution\cite{MarkovChain}. However, in our model the dimension of the parameter space is random, since it depends on the random value of $k$. Therefore, the MCMC algorithm needs to be incremented with a Reversible Jump \cite{Green95} (RJ) step, which allows to jump between models having different dimensions. 

The RJ-MCMC algorithm we propose performs $M$ sweeps  and, at each sweep, all the parameters are updated in  turn, sampling the new value of a certain parameter conditionally on the  data and all the remaining parameters  and  is able to explore models with any number of components as well as to move between models with stretched or unstretched components or any mixing of the two, thus including also Zimm and Rouse models.  At each sweep, we perform a changing dimension move to update the value of $k$ (through RJ sampling), as well as fixed dimension moves to update all the other parameters (either  through Gibbs or Metropolis-Hastings sampling \citep{Tierney94}). All the details about the algorithm and the updating moves are exhaustively explained in a dedicated paragraph of the Supplemental Material.\cite{SupplementalMaterial} 

After an initial burn-in period, during which the RJ-MCMC draws are discarded until the convergence of the chain to the target distribution is reached, the algorithm produces at each sweep $m$ a new realization
of all the parameters of the model, sampled from their joint posterior distribution. Let $(k^{(m)},  A^{(m)},\tau^{(m)},\beta^{(m)},\gamma^{(m)},\nu^{(m)})$, for $m = 1,\ldots,M$, be the sample obtained after $M$ sweeps of the algorithm. This sample  provides
the simulated joint posterior distribution of all parameters and can  be used to
estimate all the quantities of interest.

The posterior distribution of the
number of exponential components can be estimated as the proportion of times each model was visited by the
algorithm, i.e.:
$$\hat{p}(k=\ell|y)=\sum_{m=1}^M \delta_{k^{(m)},\ell}/M=M_\ell/M$$  where $M_\ell$ is the number of times the model with $\ell$ components was  visited. In practice, the model chosen will be the one with the highest posterior probability.

Once a particular model has been chosen, we
can estimate the parameters of each exponential component under that model. The estimates
of these parameters can be computed as the mean of their simulated marginal
posterior distribution, conditional on the model with $\ell$ components.  Point estimates for the weights and relaxation times of the $\ell$ exponential components are given respectively by:
$$\hat{A}_j=\sum_{m:k^{(m)}=\ell}A_j^{(m)}/M_\ell, \quad  \hat{\tau}_j=\sum_{m:k^{(m)}=\ell}\tau_j^{(m)}/M_\ell, \quad  \mbox{for }j=1,\ldots,\ell$$ 
\noindent where the sum over $m:k^{(m)}=\ell$ means over all values of $m$ in which the model with $k=\ell$ has been visited. 

Note that in case the simulated marginal posterior distribution of a certain parameter is not symmetric, other estimators than the arithmetic mean can be considered (e.g. the simulated posterior mode).

For the stretching parameters, if the simulated posterior probability of $\beta_j=1$ is larger than its prior, i.e. $\displaystyle\sum_{m:k^{(m)}=\ell}\delta_{\beta_j^{(m)},1}/M_\ell\ge 1-\zeta$, then we simply let $\hat{\beta}_j=1$, otherwise $ \displaystyle\hat{\beta}_j=\sum_{m\in S }\beta_j^{(m)}/|\mathcal S|, $ where $\mathcal S$ is the set of sweeps for which $k^{(m)}=\ell$ and $ \beta_j^{(m)}<1$, and $|\mathcal S|$ is the dimension of $\mathcal S$.

All the other model parameters do not depend on the number of exponential components  and can be estimated from their simulated marginal posterior  distributions, averaging over all the sweeps of the algorithm.

\section{Experimental Results and Analysis}\label{sec:experimental}

\subsection{PEG2000AuNP and PEG400AuNP} \label{sec:Peg}

A representative set of NSE data obtained for the PEG2000 AuNP is shown in Fig. \ref{fig:PEG2000AuNP}(a). The full set is reported in the Supplemental Material (Fig. S 2).\cite{SupplementalMaterial} As expected the nISF decays as a function of the NSE time and the decay becomes faster as \textit{Q} increases.

In the investigated \textit{Q} range we probed length scales ranging from 30 \AA\ to about 300 \AA, which include distances  on   the same  scale of the intra- and inter-polymer chains interactions, as well as distances on the scale typical of translational diffusion of the whole nanoparticle.  The translational diffusion coefficient was  estimated by DLS measurements carried out on very dilute solution of the NPs in \dwat\ ($\sim$ 0.02 weight \%). The values obtained for this parameter were $D_{LS}$= 1.35 and 3.5 \AA$^2$/ns at temperatures 280 and 318 K, respectively. From the $D_{LS}$ we can extimate the hydrodynamic radius of the PEG2000 AuNP, $R_h \sim 85$ \AA\,  through the relation $D={kT}/{6\pi \eta R_h}$ where $k_{B}$ is the Boltzmann constant, \textit{T} the temperature and $\eta$ the viscosity of \dwat \hspace{0.1cm} at the temperature \textit{T}.

\begin{figure}[tbp]
\centering
\includegraphics[height=210mm]{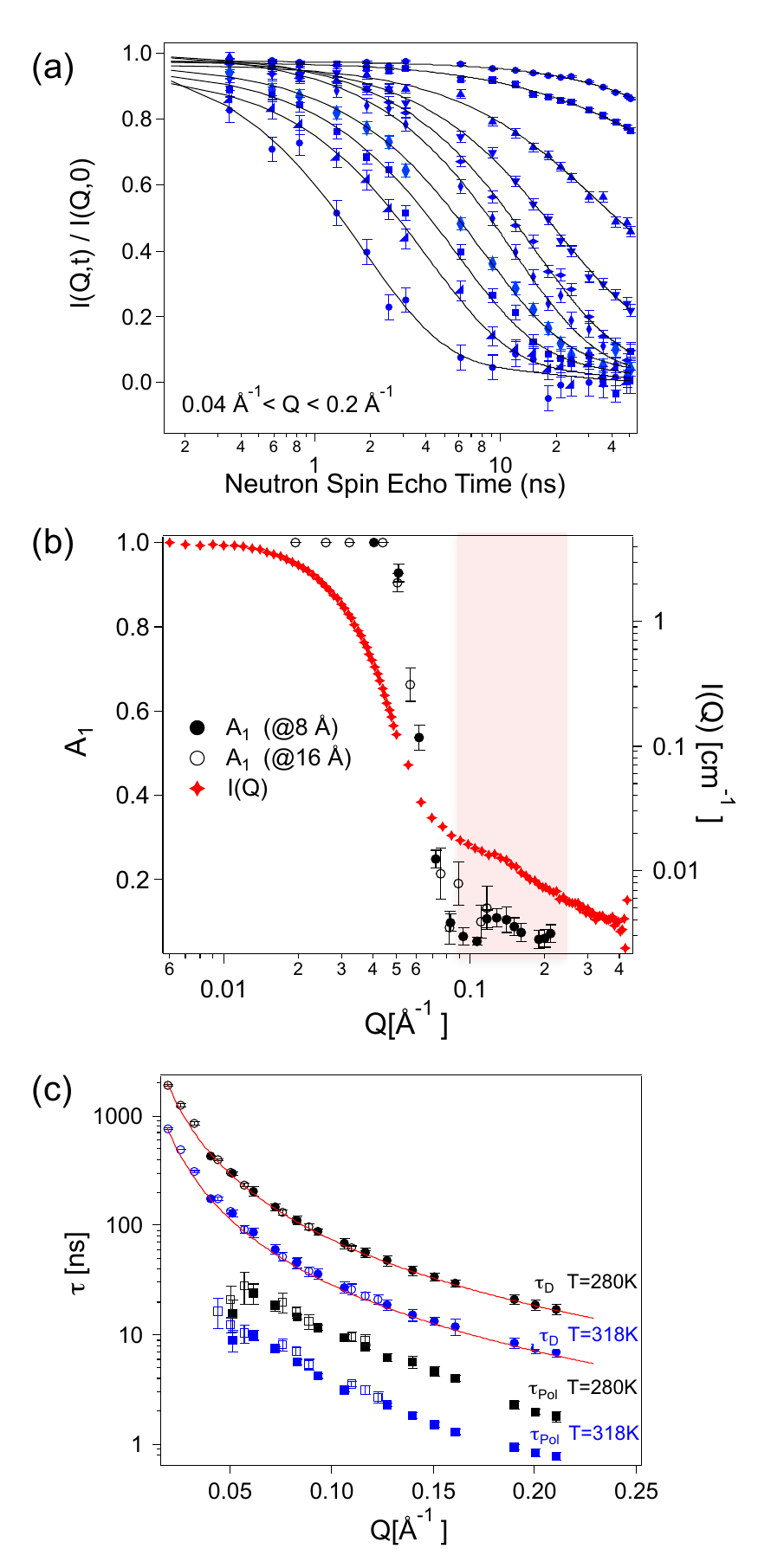}
 \caption{ (a) A representative series of NSE curves representing ${I}(Q,t)/{I}(Q,0)$ of the PEG2000 AuNPs dispersed in D$_2$O at neutron wavelength $\lambda = 10$  \AA\ and a temperature of  280 K (symbols). The $Q$ values increase from above to bottom. The lines represent the best fits to the data. (b) \textit{Q} dependence of the $A_1$ parameter obtained by the RJ-MCMC analysis and of the $I(Q)$ measured for a solution of PEG2000 AuNPs in \dwat\ at the 1\% wt. concentration; the red symbols correspond to a SANS measurement performed on the sample. (c) The two relaxation times $\tau_D$ and $\tau_{pol}$ obtained by the fitting analysis at different temperatures. The full and empty symbols correspond to the values obtained at 10 and 16 \AA.   The red lines indicate  the relaxation time obtained from the translational diffusion coefficient as derived by DLS at the respective temperatures.}\label{fig:PEG2000AuNP}
\end{figure}

At low \textit{Q}, where the condition $Q\cdot R_h \sim 1$  holds, the dynamics should be  dominated by the translational diffusion of the AuNP, whereas at higher \textit{Q} where $Q\cdot R_h >> 1$ the dynamics should be dominated by the polymer relaxations. At intermediate \textit{Q} values, a $Q$ dependent combination of these two contribution is expected. With this in mind and in analogy with Arriaga {\it et al.} \cite{arriaga2009}  we initially described the decay of ${I}(Q,t)/{I}(Q,0)$ as:   
\begin{equation} \label{eq:fitfunction}
{I}(Q,t)/{I}(Q,0)=A(Q)\cdot F_{NP}(Q,t)+(1-A(Q))\cdot F_{Pol}(Q,t)
\end{equation}
where 
\begin{equation}\label{eq:TransDiff}
F_{NP}(Q,t)=\exp(-t/\tau_D)
\end{equation}
is the translational diffusion of the NP, and
\begin{equation}
 F_{pol}(Q,t)=\exp\left[(-t/\tau_{pol})^{\beta}\right]
 \end{equation}
is the term approximately associated to the polymer dynamics (see above)  often described with a Kohlrausch-Williams-Watts function. In the previous equations $A$ is the $Q$-dependent relative portion of translational diffusion with respect to the polymer dynamics contribution, $\tau_D=1/DQ^2$ is the characteristic time of the translational diffusion of the nanoparticle in D$_2$O,  $\tau_{pol}$ is a characteristic time associated to the polymer dynamics (although it contains also  the translational diffusive dynamics (c.f. Eq. (\ref{IH_series})), and $\beta$ the stretching exponent, which  often accounts for the possibility of having a distribution of relaxation processes. However, we wanted to leave open the possibility to  describe the polymer relaxations with  a more general model. Consequently, we chose to describe the polymer relaxations with  sums of exponential (stretched or simple) functions in order to find the combination that described the data with the greatest statistical significance. 

In Eq. (\ref{IH_series2}) we imposed  the first component to be  a simple exponential (which corresponds, as discussed in Sec.\ref{sec:bayes}, to set $P(\beta_1=1)=1$ as degenerate prior distribution for $\beta_1$), since it is associated to the translational diffusion of the NP (see for example Eq. (\ref{IH_series})). As already discussed in Sec.\ref{sec:bayes}, we assumed $D\sim \mathcal{N}(\mu_D, \sigma^2_D)$. We allowed the remaining components to  either be simple or stretched exponentials. 

The results of the fitting procedures are listed in Tables S I, S II, S III and S IV in the Supplemental Material.\cite{SupplementalMaterial}  The column labelled $P(k|y)$ lists the probability associated to the most statistically supported model conditional to our data. In Fig. \ref{fig:PEG2000AuNP}  we show a summary of the main results concerning the fitting analysis for the PEG2000 AuNP. 
For \textit{Q} $<$ 0.04 \AA$^{-1}$ the RJ-MCMC algorithm privileged a solution with only the translational diffusive component at both measured temperatures (Fig. \ref{fig:PEG2000AuNP}(b) and \ref{fig:PEG2000AuNP}(c)). 

Above this value of \textit{Q}, the solution with two relaxations was always  chosen with a probability $P(k=2|y)$ usually between 70 and  90\% especially at high \textit{Q} and at the higher temperature. The mean $\beta_2$ stretching coefficient was estimated by the algorithm to be  close to 1 and in any case the probability to have $\beta_2=1$ was most of the times  greater than 90\%.   

In Fig. \ref{fig:PEG2000AuNP}(b)  the trend in \textit{Q} of the parameter $A(Q)$ is displayed at the temperature of 280 K. At  \textit{Q} $<$ 0.04 \AA$^{-1}$  the main contribution to the dynamics came from the translational diffusion of the NP. At $Q$ $>$ 0.1 \AA$^{-1}$ the dominant contribution was the polymer dynamics. In the intermediate \textit{Q} regime 0.04 \AA$^{-1}$ $<$ \textit{Q} $<$ 0.1 \AA$^{-1}$, we observed a transition between these two regimes.

The value of the translational diffusion coefficient estimated by the algorithm was in perfect agreement with the one measured by DLS (see Fig. \ref{fig:PEG2000AuNP}(c), solid red line) which confirms that the contribution of the rotational term can actually be neglected. From Fig. \ref{fig:PEG2000AuNP}(c) it is also clear that the  relaxation times estimated  at the two  wavelengths are  consistent with each other. As expected, the relaxation times decreased both with increasing the temperature and \textit{Q}. 
 
\begin{figure}[h]
	\centering
	\subfigure[]{\includegraphics[width=0.4\linewidth]{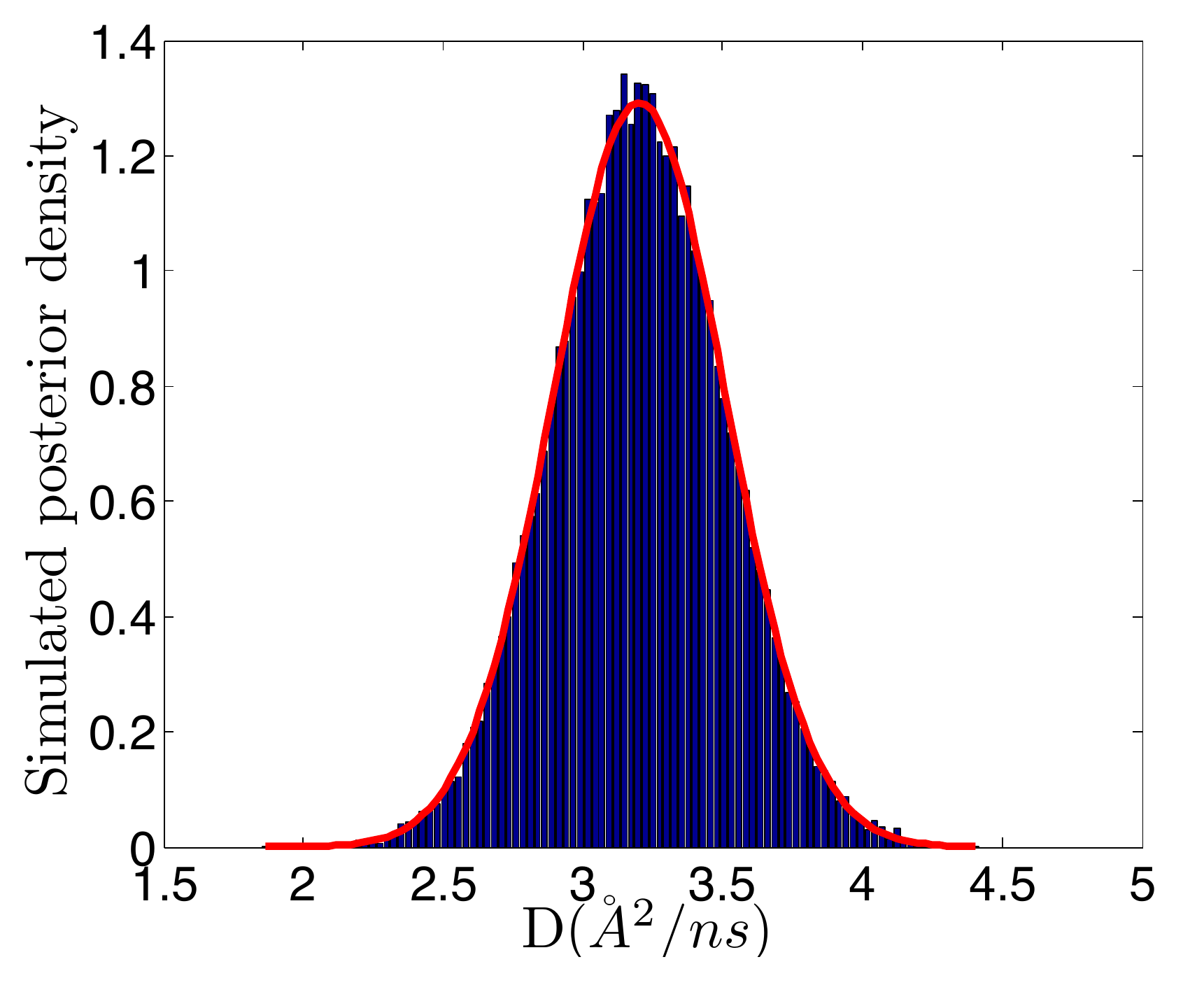}}	
	\subfigure[]{\includegraphics[width=0.36\linewidth]{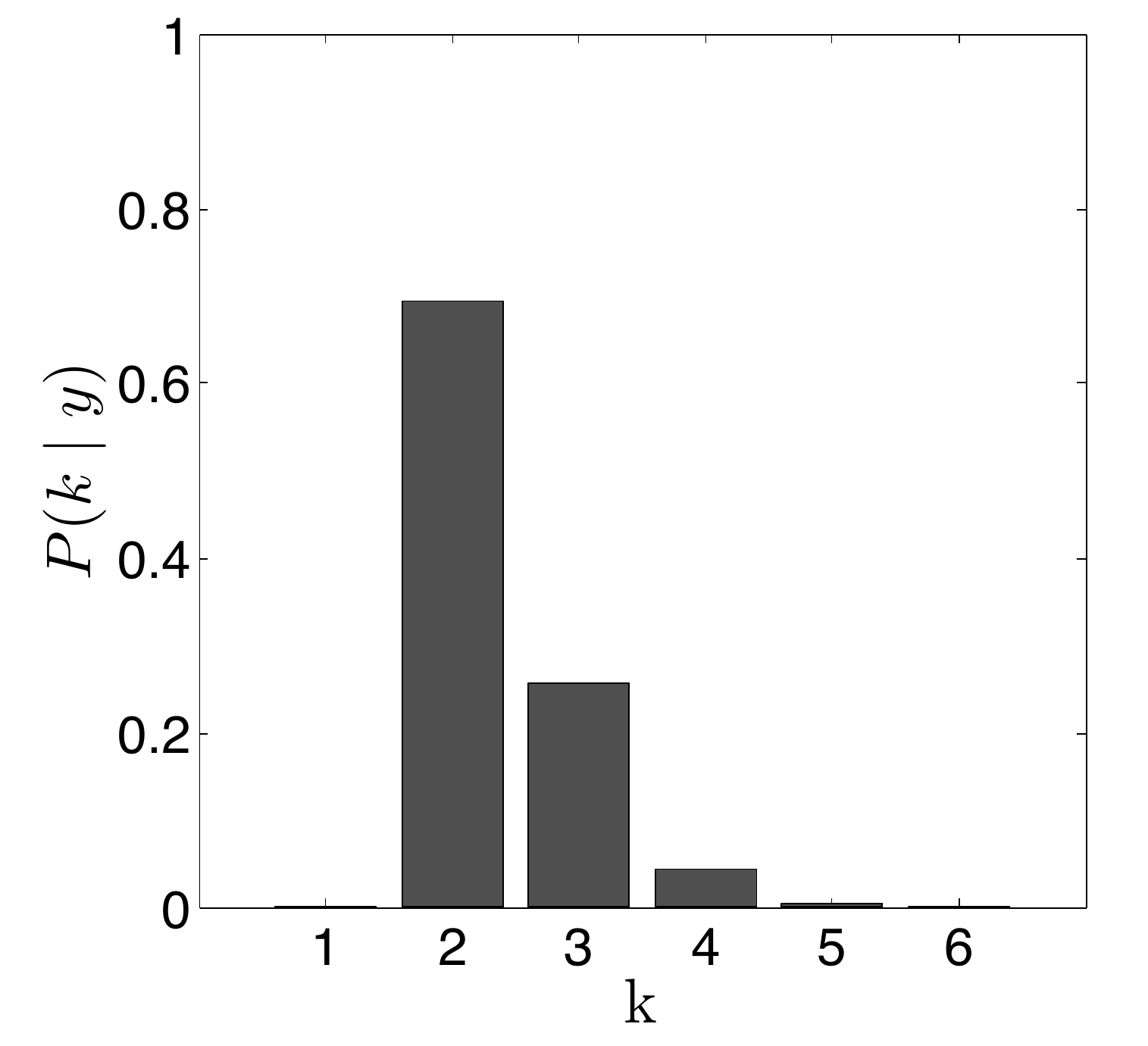}}	\\		
	\subfigure[]{\includegraphics[width=0.4\linewidth]{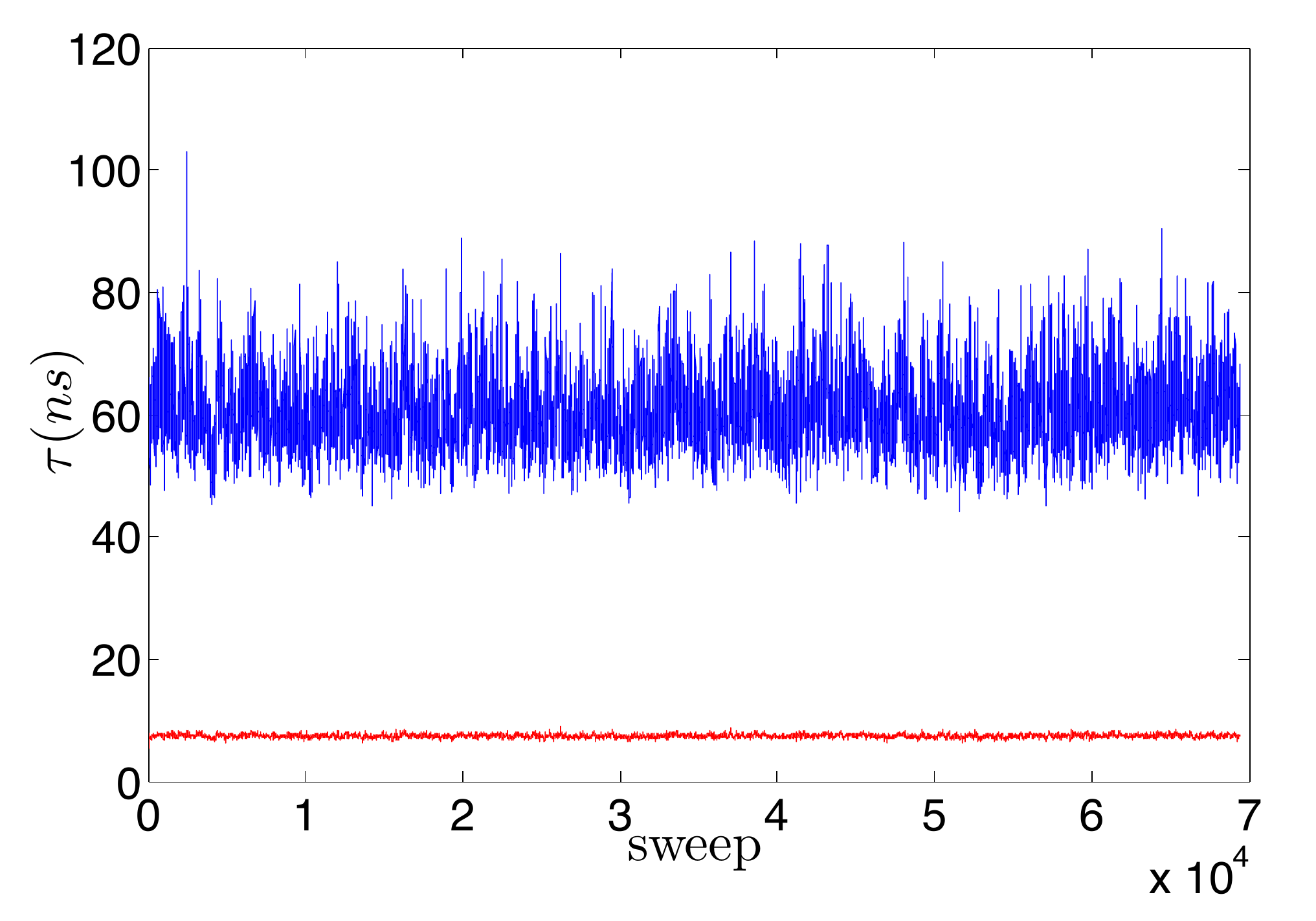}}
	\subfigure[]{\includegraphics[width=0.4\linewidth]{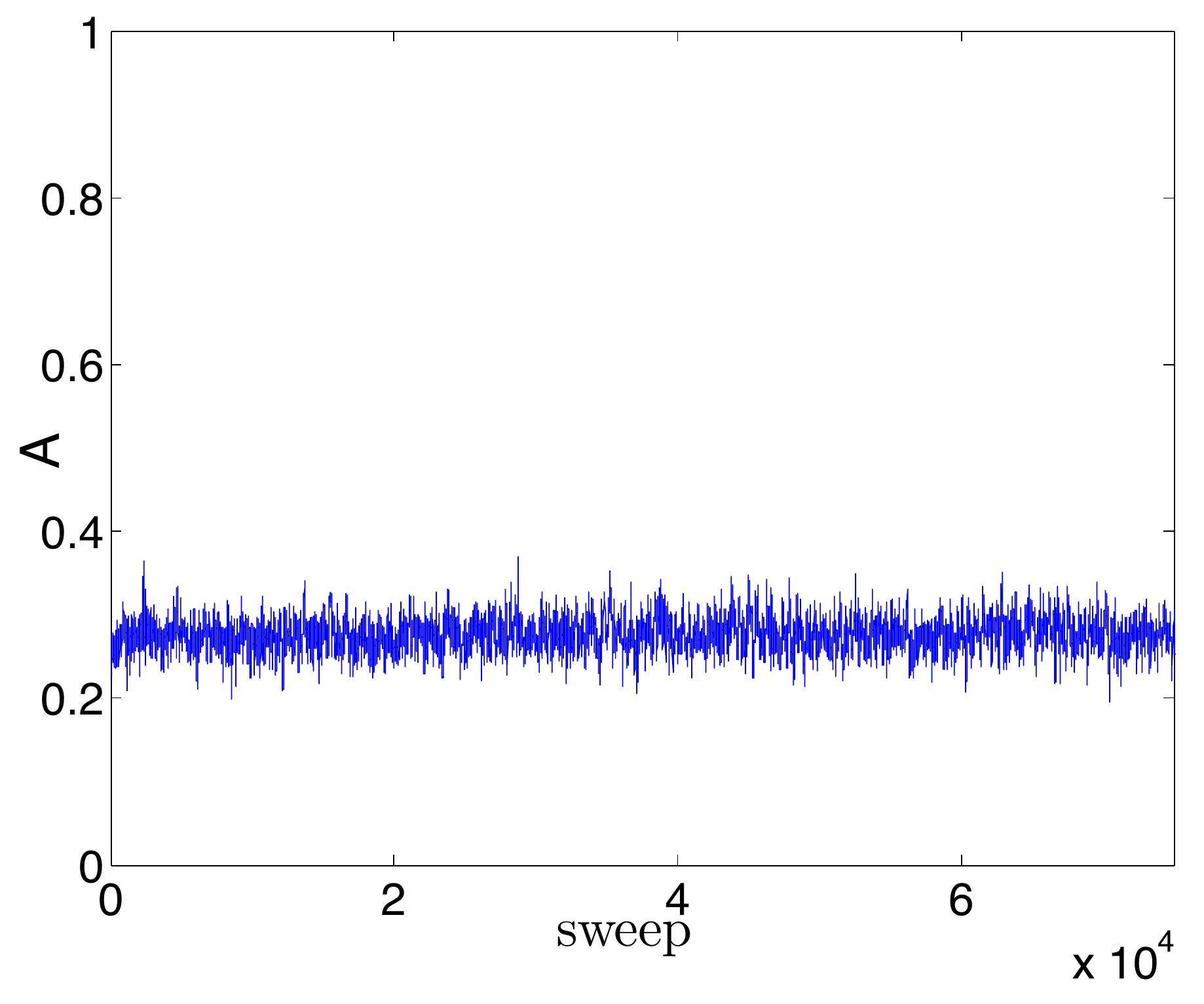}}
	\caption{(a) An example of the posterior distribution for the parameters $D$ for the PEG2000AuNPs at 318 K, $\lambda=10$ \AA, $Q=0.72$ \AA$^{-1}$. The solid red curve is a Gaussian fit to the simulated posterior distribution; (b) Posterior distribution for the number of relaxation components for the same example in (a);  (c)  Trace Plot for the relaxation time $\tau_{D}$ (bottom curve) and $\tau_{pol}$ (upper curve);  (d) Trace plot for the $A_1$ parameter.}	\label{fig:fewresults}
\end{figure}

\begin{figure}[tbp]
\centering
\includegraphics[height=100mm]{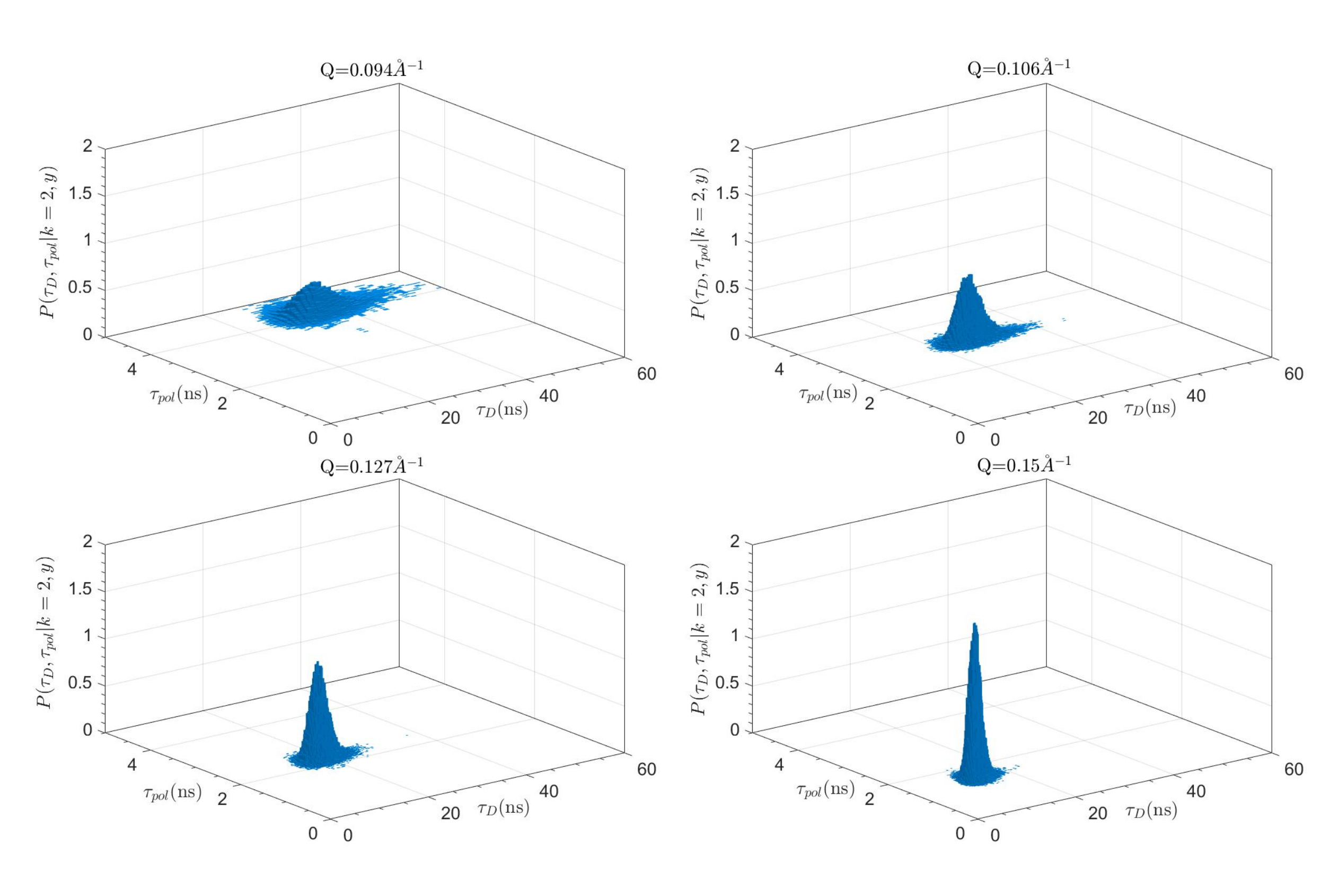}
 \caption{Joint posterior distribution at four selected $Q$ values, of the two relaxation times, conditional on k=2 and marginalized with respect to the other model parameters for the PEG2000 AuNP sample at $T$=318 K and $\lambda$=10 \AA.}\label{fig:Fig3Dposterior}
\end{figure}

 In Fig. \ref{fig:fewresults} we report as  example  a typical set of results coming from the fitting algorithm obtained at $\lambda=10$ \AA\ and at $Q=0.72$ \AA$^{-1}$. 
 Figure \ref{fig:fewresults}(a) shows an example of the posterior distribution of the translational diffusion coefficient $D$  at $Q=0.72$ \AA$^{-1}$.  In this specific case the simulated posterior distribution was a Gaussian, as shown by the fit (red curve). The spread of this distribution is related to the polydispersivity of the NP and to the errors bars of the experimental $I(Q,t)$. Figure  \ref{fig:fewresults}(b) displays the posterior distribution of the number of components in Eq. (\ref{eq:datamodel}) probed by the algorithm. In Fig. \ref{fig:fewresults}(c) and \ref{fig:fewresults}(d)  the Trace Plots for the two relaxation times  $\tau_D$ and $\tau_{pol}$ and for the \textit{A} parameter are shown,  always for the same $\lambda$ and $Q$.  In Fig. \ref{fig:Fig3Dposterior} we report as an example the joint posterior distribution $P(\tau_D, \tau_{pol} \vert k=2,y)$ for the two relaxation times marginalized to all the other model parameters  as a function of $Q$ for the PEG2000AuNP sample at 318 K. It can be noticed that at small $Q$ values, the posterior distribution of $\tau_D$ is more dispersed than that of $\tau_{pol}$, implying a minor accuracy in the estimation of the first parameter, compared to the second one. As $Q$ increases, the difference	between the dispersions of the two parameters vanishes and, in addition, the joint 	distribution becomes more concentrated, implying that more precise estimates of both parameters can be obtained at large $Q$ values. Also notice the absence of correlation between the two parameters.
 
\begin{figure}[tbp]
\centering
\includegraphics[height=210mm]{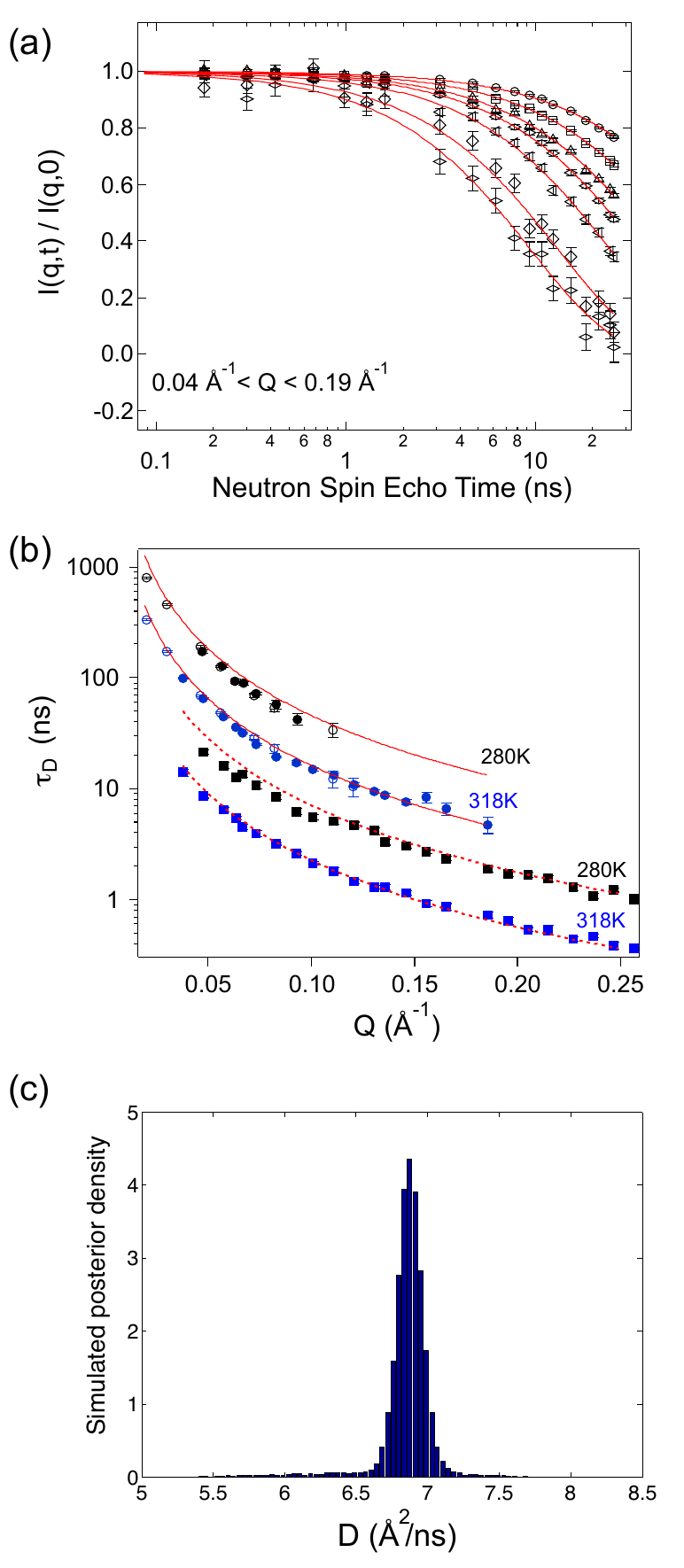}
 \caption{(a) Representative  set of  NSE curves  measured on the PEG400 AuNPs at  $\lambda = 8$ \AA\ and \textit{T}=318 K;  (b) Diffusion relaxation time as derived by the RJ-MCMC algorithm  for the PEG400 AuNPs (circles)  and  for the PEG400 polymers in 5\% solution in \dwat \ (squares).  The empty and solid symbols refer to $\lambda= 8$ and 16 \AA\, respectively. The continuous red lines correspond to the relaxation times obtained from the translational diffusion coefficients  as measured by DLS measurements  for the PEG400AuNPs.  The dashed lines indicate the relaxation times obtained from the translational diffusion coefficient of PEG400 solution as obtained from literature\cite{waggoner1995}; (c) An example of the simulated posterior distribution for the diffusion coefficient obtained at \textit{T} = 318 K and  $Q=0.048 $ \AA$^{-1}$.}\label{fig:FigurePEG400AuNP}
\end{figure}

 A representative set of the NSE data measured on the PEG400 AuNPs dispersed in \dwat\ are shown in Fig. \ref{fig:FigurePEG400AuNP}(a).  At both the investigated temperatures the most probable model predicted by the algorithm to describe the decay of ${I}(Q,t)/{I}(Q,0)$ was a single exponential relaxation. At 280 K, even if the quality of data at $Q\ge0.8$ \AA$^{-1}$ degraded (see Fig. S4 in the Supplemental Material\cite{SupplementalMaterial}), the probability assigned by the algorithm  to a one-component model was, except for a few datasets, significantly greater than 60\%. The estimated relaxation time at each $Q$ value is very close to the one expected for the NP translational diffusion as measured by DLS ($D_{LS} = 2.2$ \AA$^2$/ns; cf. Fig. \ref{fig:FigurePEG400AuNP}(b), red solid lines).  At 318 K the situation was clearer in view of an improved data quality. This allowed an unambiguous fitting of the relaxation curves given that the algorithm assigned  to a one-component relaxation model  a posterior probability  $P(k=1|y)$  almost always above 90\%. The translational diffusion coefficient from the fits is in agreement within the errors with that obtained with the DLS ($D_{LS} = 6.2$ \AA$^2$/ns) and corresponds to a hydrodynamic radius of $\sim$ 48 \AA. The complete set of  parameters obtained by the fits are listed in Tables S V, S VI, S VII and S VIII in the Supplemental Material.\cite{SupplementalMaterial}

\subsection{Polymer solutions of PEG2000 and PEG400}

\subsubsection{PEG2000}  \label{sec:PEG2000results}

\begin{figure}[tbp]
\centering
\includegraphics[height=140mm]{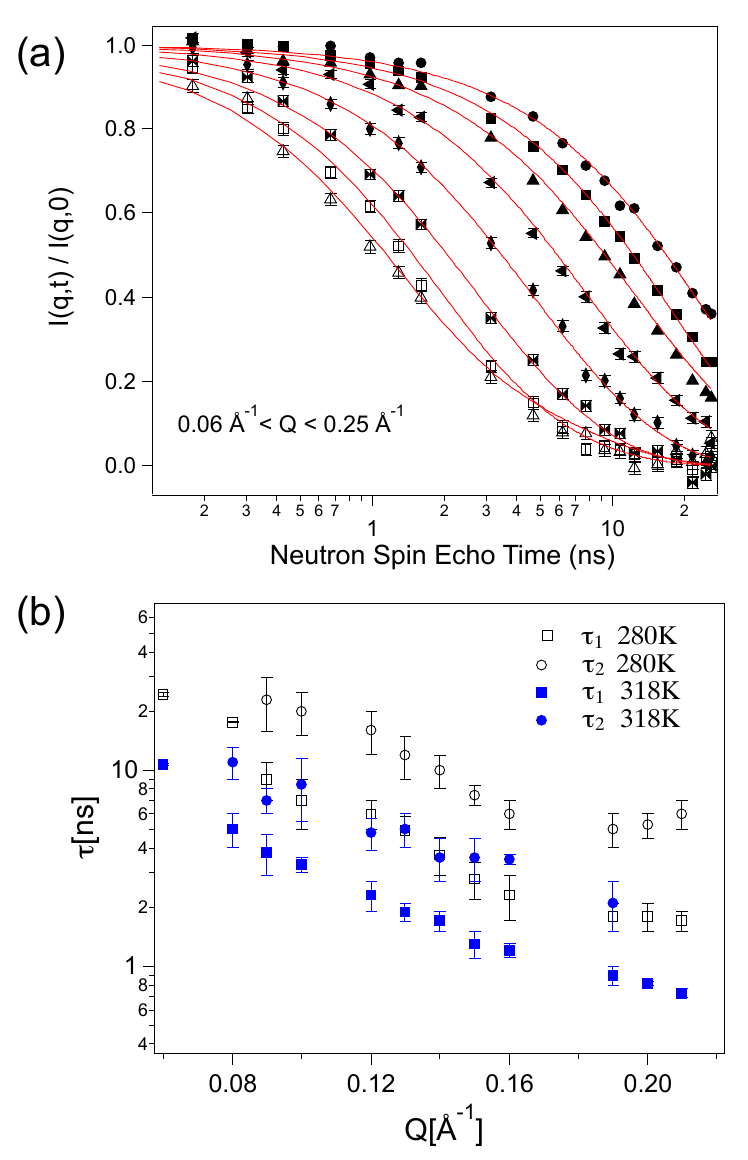}
 \caption{(a) Representative  set of  NSE curves  measured on the PEG2000  at  $\lambda = 8$ \AA\ and \textit{T}=280 K;  (b) The two relaxation times $\tau_1$ and $\tau_{2}$ obtained by the fitting analysis at different temperatures. }\label{fig:FigPEG2000}
\end{figure}

In order to better characterise  the  dynamics of the polymer grafted to the NP, we compared the results obtained with the PEG coated NPs and  those obtained with a reference system identified as a solution of PEG homopolymers. We used PEG2000 and PEG400 that possess a molecular weight very similar  to those tethered to the NPs. 
 As previously shown \cite{maccarini2010},  the concentration of polymer as a function of the distance from the gold surface is not uniform, but varies being rather high in the vicinity of the surface of the NP and lower away from it. Ideally, we should compare the results obtained on the PEG coated NP with several PEG solutions at different concentrations. The PEG2000 was studied in \dwat \hspace{1pt} at weight concentrations of 10\% 20\% and 45\% (the latter is the solubility limit for PEG2000 in \dwat)  as a reference.  However, due to an unfavorable combination of coherent and incoherent components that contribute to the neutron polarization with opposite sign, the measurements at the higher concentrations (20 and 45\% in weight) suffered of a too weak spin echo signal and were not taken into account.   The choice of the reference system was based on a compromise between the counting statistics and the available measuring time. Ultimately, the PEG2000 and PEG400 were studied at concentrations of 10\% and   5\% in weight, respectively.  A representative set of data obtained for the PEG2000 at 10\% is displayed in Fig. \ref{fig:FigPEG2000}(a). 
 
 The protocol used to analyse the NSE data of the polymers was slightly different from the one used	 for the NPs, as we did not use any informative prior for the first relaxation. In fact, we left the algorithm free to search for the most statistically probable solution. The prior distribution of the relaxation times was chosen as to be uniform in an interval $0\div\tau_{\max}$.  

In a first attempt, we used a combination of simple exponential functions to describe  ${I}(Q,t)/{I}(Q,0)$. The results for  PEG2000   homopolymers  at a concentration of 10\% and at $\lambda=8$ \AA\ are shown in Fig. \ref{fig:FigPEG2000}(b). The parameters obtained from the fits are listed in Tables S IX and S X in the supporting information.

\begin{figure}[htbp]
	\centering
	\includegraphics[height=100mm]{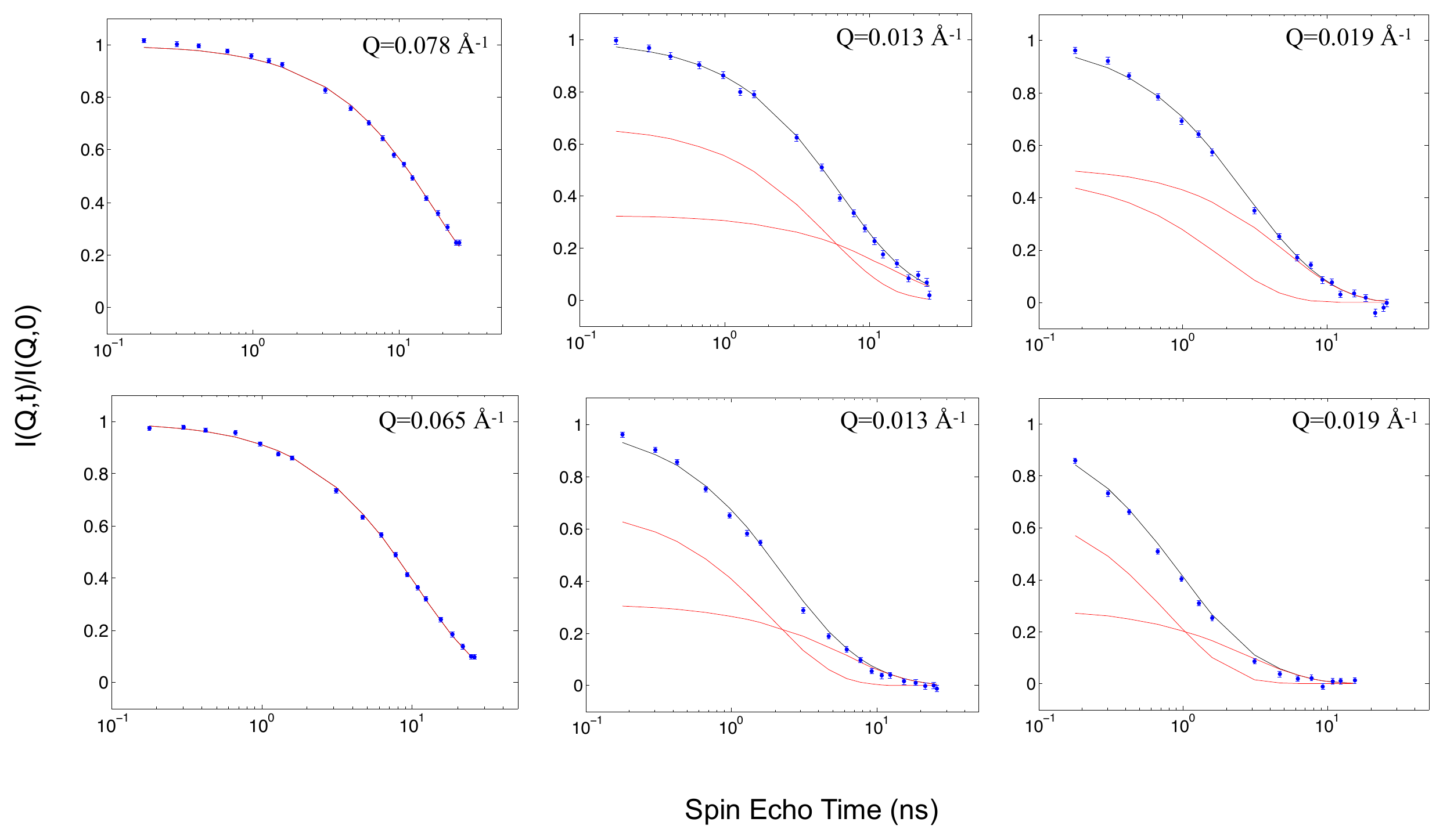}
	\caption{NSE curves representing ${I}(Q,t)/{I}(Q,0)$ measured on the PEG2000 homopolymer 10\% concentration, at 280 K (top panels) and 318 K (bottom panels) at three selected $Q$ values. The best fit (black line) and the relaxation components (red curves) are shown. }	\label{fig:fitresultsPEG2000}
\end{figure}

At \textit{T}=280 K, the algorithm favors a single exponential for  $Q\leq0.08$ \AA$^{-1}$ and two exponentials for $Q >0.08$ \AA$^{-1}$ (see Fig.  \ref{fig:FigPEG2000}(b) and Fig. \ref{fig:fitresultsPEG2000} top panels). At \textit{T}=318 K we have a similar behavior up to  $Q=0.19$ \AA$^{-1}$.

Above this value, the algorithm would still favor the solution with two components. However, while the parameters of the first component are nicely estimated to continue the trend of their respective values at lower $Q$s, the parameter $\tau_2$ results scarcely identified. In fact, the simulated posterior distribution of $\tau_{2}$ turns out to be substantially uniform in a large range of time (Fig. S 5(b)). As a consequence, any value of $\tau_{2}$ sufficiently large would define a second exponential function able to fit the decay of the time correlation function at large Spin Echo Time.  We, therefore, decided to retain both components in the model (a single component would simply misconsider the last few points; see Fig. S 5(a) in the Supplemental Material\cite{SupplementalMaterial}) but we do not trust estimates for the slower dynamics, since meaningless. The algorithm still provides the best probabilistically supported solution given the experimental data, but at the highest \textit{Q} values the number of data points available for a precise determination of $\tau_{2}$ is insufficient.  
The application of our method particularly for datasets as the ones at $Q>0.19$ illustrates the efficacy of the Bayesian analysis since we can infer the reliability of the parameter estimates from the simulated marginal posterior distributions of the different parameters.  This represents effectively a diagnosis tool to exploit at maximum the collected experimental data.

\begin{figure}[htpb]
 	\centering
 	\includegraphics[height=160mm]{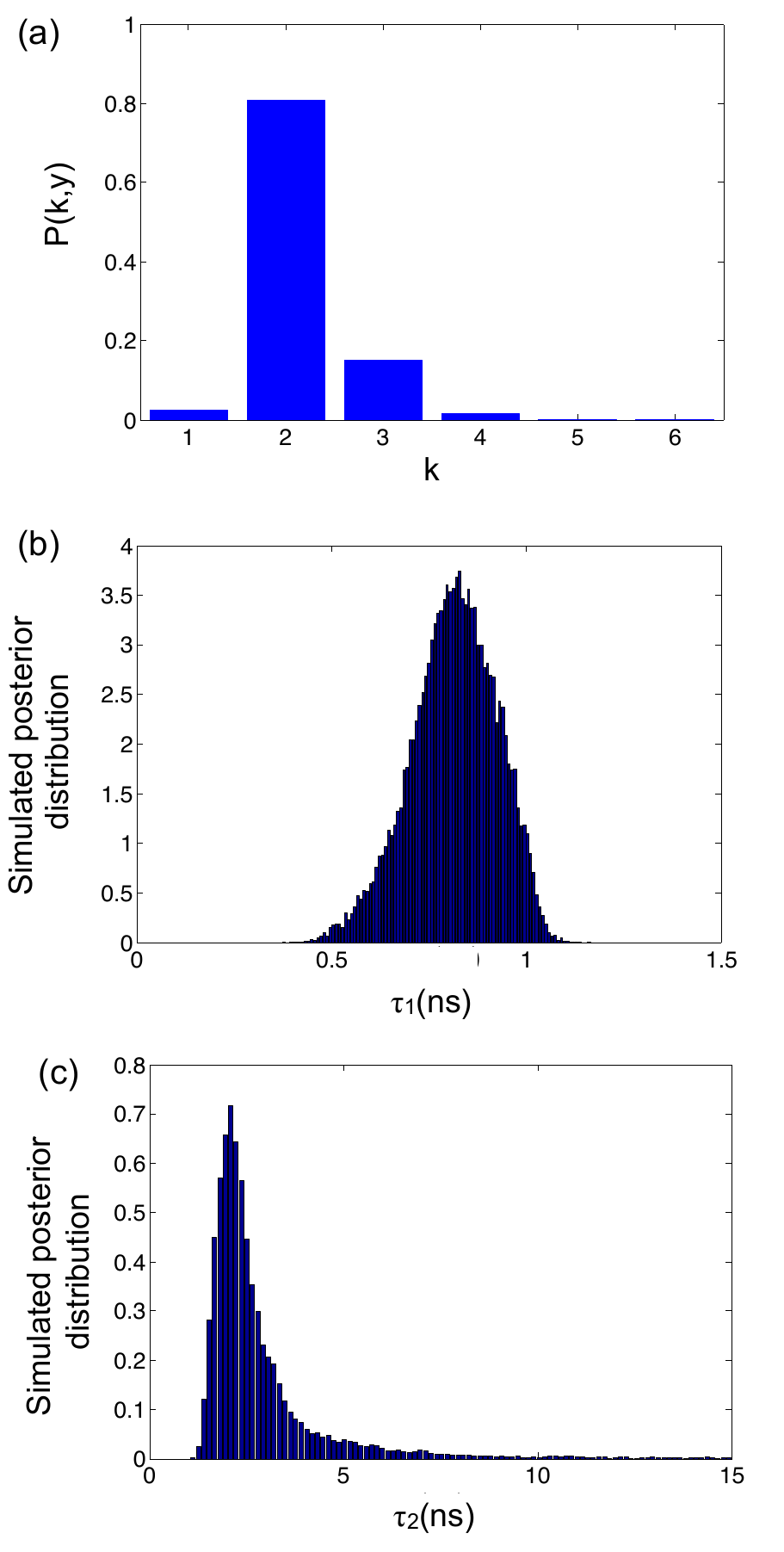}
 	\caption{We provide as example the posterior distribution obtained for PEG2000 homopolymer at 280 K and at $Q=0.19$ \AA$^{-1}$  for  (a)  the number $k$ of relaxation components (b)   $\tau_{1}$ and  (c) $\tau_{2}$.}	\label{fig:fitresultsPEG2000II}
 \end{figure}

 We stress that the posterior probability for the chosen values of $k$ is always larger or much larger than 60\% (see an example in Fig. \ref{fig:fitresultsPEG2000II}). Also,  the estimate of the relaxation times is accurate for the majority of data sets and  providing for the posterior parameter distribution function a Gaussian distribution or at least a well-shaped unimodal distribution although not necessarily symmetric (see Fig. \ref{fig:fitresultsPEG2000II} panel b) and c)).

\begin{figure}[tbp]
\centering
\includegraphics[height=70mm]{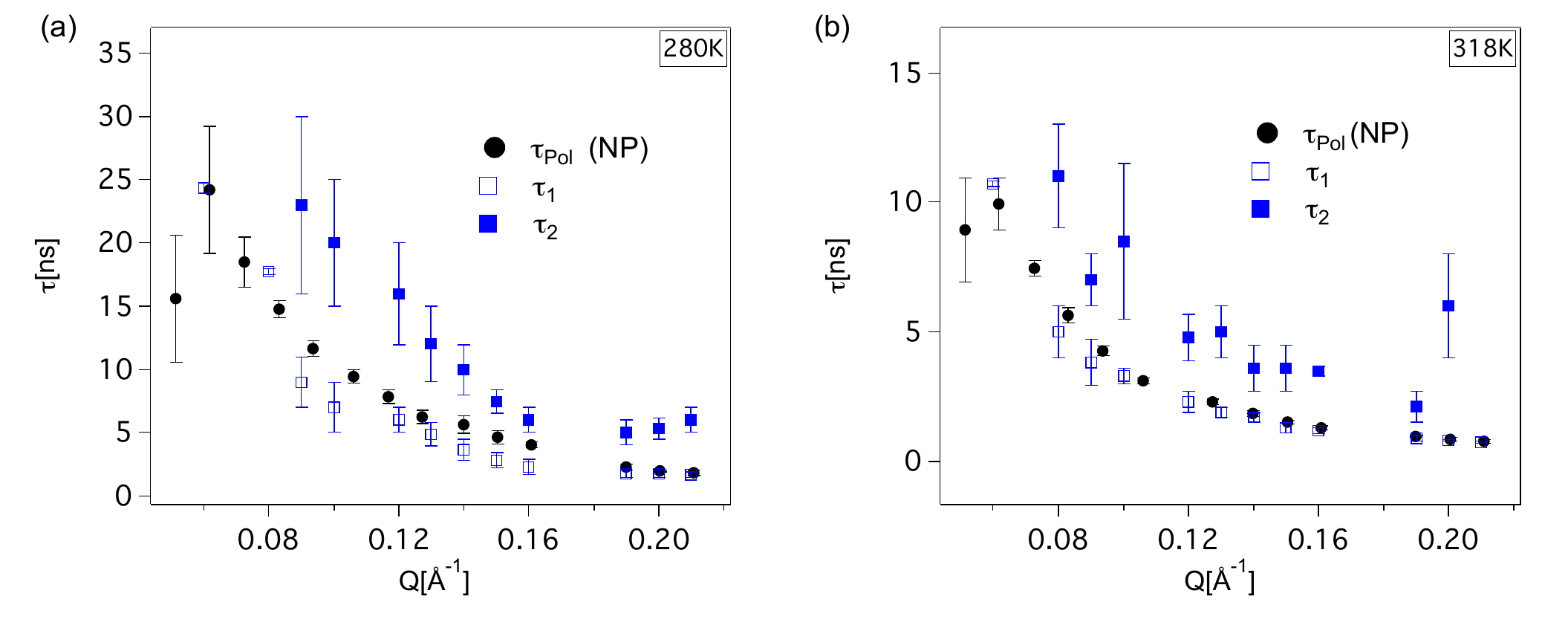}
 \caption{Characteristic relaxation times $\tau_1$ and $\tau_2$ for the homopolymer PEG2000 solution at 10\% concentration and relaxation time of the polymer in the AuNP, $\tau_{pol}$, at (a) \textit{T}=280 K and (b) \textit{T}=318 K.}\label{fig:Figure-TauComparison}
\end{figure}

\begin{figure}[htpb]
	\centering
	\includegraphics[height=100mm]{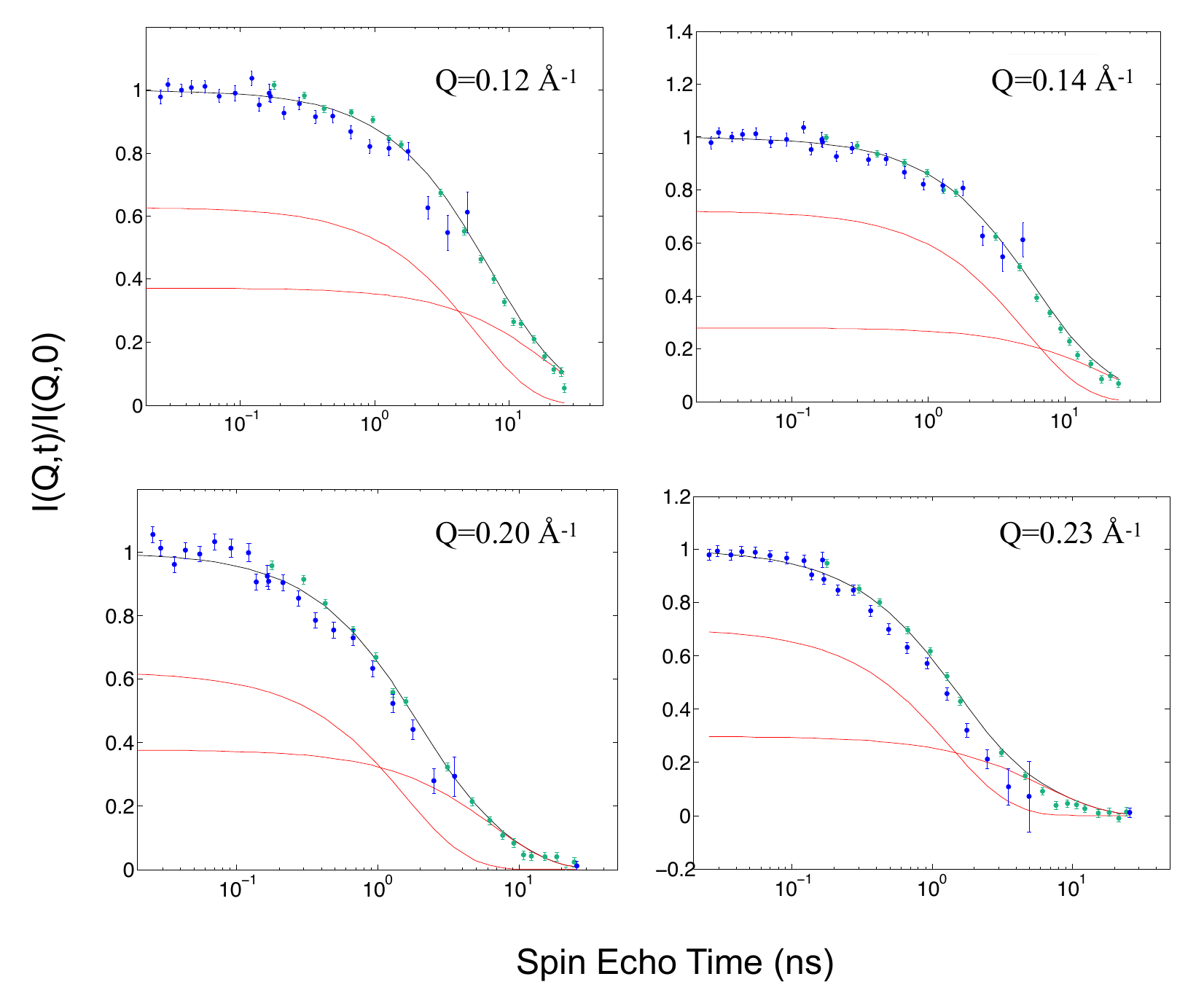}
	\caption{NSE curves for the data collected with IN11 (blue symbols) and IN15 (green symbols) joined together on a PEG2000 homopolymer solution in \dwat\ at 10\% concentration, at four selected $Q$ values and \textit{T}=280 K. }	\label{fig:spettroni}
\end{figure}

Further measurements on the PEG2000 homopolymers dissolved in \dwat\  at  10\% in weight  were carried out on the spin echo spectrometers IN11 at ILL with an incident wavelength of $8.5$ \AA. These measurements expanded the accessible time and momentum transfer, covering a time window between 0.02 and 5 ns and a \textit{Q} range between 0.09 and 0.42 \AA$^{-1}$. In order to further validate the scenario so far achieved we joined the data collected with the two spin echo instruments IN11 and IN15 extending the bayesian MCMC data analysis to this enlarged time window. The available data from IN11 were only at 280 K. In Fig. \ref{fig:spettroni} we show the relaxation curves at four selected momentum transfer wavevectors ($Q$=0.12, 0.14, 0.20 and 0.23 \AA$^{-1}$) at which we collected the data from both spectrometers. 

At $Q$=0.12, 0.20 and 0.23 \AA$^{-1}$ we re-obtain substantially the same results we found from the analysis of the IN15 data only, both for the characteristic relaxation times and the weights \textit{$A_1$} and (1-$A_1$). At \textit{Q}=0.14 \AA$^{-1}$ there is a slight difference for the faster relaxation, but the results are fully consistent within the uncertainties (Tab. \ref{tab:confrontospettroni}).

\begin{table}
\centering
\caption{Comparison between the values  of the relaxation times and their weights obtained by  fitting only the IN15 data and by fitting the nISF obtained by gathering together the data from IN11 and IN15.}
\footnotesize
\begin{tabular}{|c|c|c|c|c|c|} 
\hline
      Q [\AA$^{-1}$]  &    Instrument     & $ \tau_1$ [ns]    & $ \tau_2$ [ns]     &   $ A_{1}$  \\ \hline \hline
      0.12                  &   IN15               &         6$\pm$1   &       16$\pm$4       &            0.59                           \\ 
                              &   IN15 \& IN11   &    6$\pm$1     &       12$\pm$ 4    &            0.60                            \\ \hline
      0.14                 &   IN15                &   3.7$\pm$ 0.8   &       10 $\pm$ 2  &            0.46                             \\ 
                              &   IN15 \& IN11   &   4.7 $\pm$ 0.8  &     10$\pm$ 4   &            0.60                          \\ \hline
      0.20                 &   IN15                &   1.8$\pm$ 0.3   &    5.3 $\pm$ 0.8  &            0.49                             \\ 
                              &   IN15 \& IN11   &   1.6 $\pm$ 0.2  &    5.2$\pm$ 0.8   &            0.62                           \\ \hline 
      0.23                 &   IN15                &   1.6$\pm$ 0.2   &    5 $\pm$ 1        &            0.69                            \\ 
                              &   IN15 \& IN11   &   1.3 $\pm$ 0.2  &    4.0$\pm$ 0.8   &            0.69                            \\ \hline 
                      
  \end{tabular}
    \label{tab:confrontospettroni} 
\end{table}

 \subsubsection{PEG400}

The study of the PEG400 homopolymer solution has been undertaken at a concentration of 5\% wt. The description once again is rather simple. As in the case of the PEG400AuNP's, one relaxation is sufficient to give account of the time correlation decays at both the investigated temperatures. The posterior distribution function for $k$ predicts a probability $P(k=1\mid y)$ for a single relaxation model always in the range $80\div 98\%$ for all the spectra analysed, with few exceptions where anyway such probability was larger than $60\%$.
The relaxation times obtained at the two temperatures are summarised in Fig.  \ref{fig:FigurePEG400AuNP}(b). These relaxation times are coherent with the values of translational diffusion of PEG400 at the same temperature and concentration (cf. Fig. \ref{fig:FigurePEG400AuNP}(b), dot red lines). \cite{waggoner1995} A selection of the experimental NSE  curves at the two investigated temperatures and at three selected $Q$ values and fits are displayed in Fig. S 6 of the Supplemental Material.\cite{SupplementalMaterial} In Fig. S 7 we report two examples of the simulated posterior distribution function for the relaxation time $\tau_{1}$ for two different values of $Q$ at \textit{T}=318 K. In these  cases  the posterior distribution functions look unimodal and almost symmetric. The relaxation times obtained for the PEG400  indicate a  faster dynamics compared to that found in the PEG400 AuNP (Fig.  \ref{fig:FigurePEG400AuNP}b).

\section{Discussion}

The fitting analysis of the PEG2000 AuNP data details two relaxation processes that occur in the time window probed by the NSE experiment. The slower relaxation process with characteristic time $\tau_D$  corresponds to the translational diffusion of the PEG2000 AuNP. As it appears from Fig. \ref{fig:PEG2000AuNP}(c), it perfectly overlaps with the relaxation times obtained  from the expression $\tau_D=1/DQ^2$ where $D$ is the translational diffusion as derived by DLS measurement of PEG2000 AuNP at infinite dilution and at the respective temperatures.  To further check if the interaction of the NP due to the finite concentration used in the experiments had a significant effect on the dynamics, we obtained the structure factor $S(Q)$ by performing SANS experiments at increasing concentrations (see Fig. S 3  in the Supplemental Material\cite{SupplementalMaterial}). Following the de Gennes narrowing, a significant interaction would produce a modulation of the translational diffusion according to the inverse of the structure factor $1/S(Q)$. As it is shown in Fig. S 3 this modulation is barely visible within the error bars. This makes the assumption of negligible interactions between the nanoparticles  reasonably applicable in this case. 

While at small $Q$ the signal is dominated by the coherent contribution due to the translational diffusion of the PEG AuNP, at high $Q$ the signal is dominated by the chain relaxation ($A_1\le0.2$)  and it is composed by similar proportion of coherent and incoherent (Fig. S1).  The observation of only one single exponential decay at those high $Q$-values implies that the coherent and incoherent scattering have a similar relaxation behavior within the limit of the sensitivity of the technique. A possible explanation for this feature is that we are looking to a system of polymers closely tethered to a small surface from the perspective of a relatively large length-scales ($> 25 $ \AA). In these condition it is reasonable to think that the polymers cannot move independently from each other and that the self and coordinated motion of the polymers relax on similar timescales.

The second relaxation process is related to the dynamics of the polymer chains tethered to the NP,  as shown by the fact the $\tau_{pol}$  resembled the relaxation times found in solution of PEG2000 in \dwat\ (Fig. \ref{fig:Figure-TauComparison}). The relative contribution of the two relaxations processes is expressed by the $Q$-dependent parameter $A_{1}$, that represents the portion of translational diffusion,  with the sum rule $A_{1}+A_{2}=1$. As shown in Fig. \ref{fig:PEG2000AuNP}(b),  the translational diffusion of the NP dominates at low $Q$ and the polymer dynamics dominates at high $Q$. Interestingly,  the  trend of \textit{$A_{1}$} as a function of \textit{Q} suggests a maximum at the same \textit{Q} where the $I(Q)$ obtained by SANS measurements presents a shoulder (see the shaded region in Fig. \ref{fig:PEG2000AuNP}(b)). In fact the form of the function $A_{1}(Q)$ reflects a modulation of the structure factor. At the $Q$ values where the maximum occurs, the incoherent part of the scattering is around 30\%. This incoherent part also contribute to the $A_{1}(Q)$ with a term given by $I_{inc}(Q,\infty)/I_{inc}(Q,0)$. This term is the usually called EISF (Elastic Incoherent Structure Factor), which relates to the volume where the hydrogen atoms are confined in, i.e. the nanoparticle corona.

\begin{figure}[tbp]
\centering
\includegraphics[height=75mm]{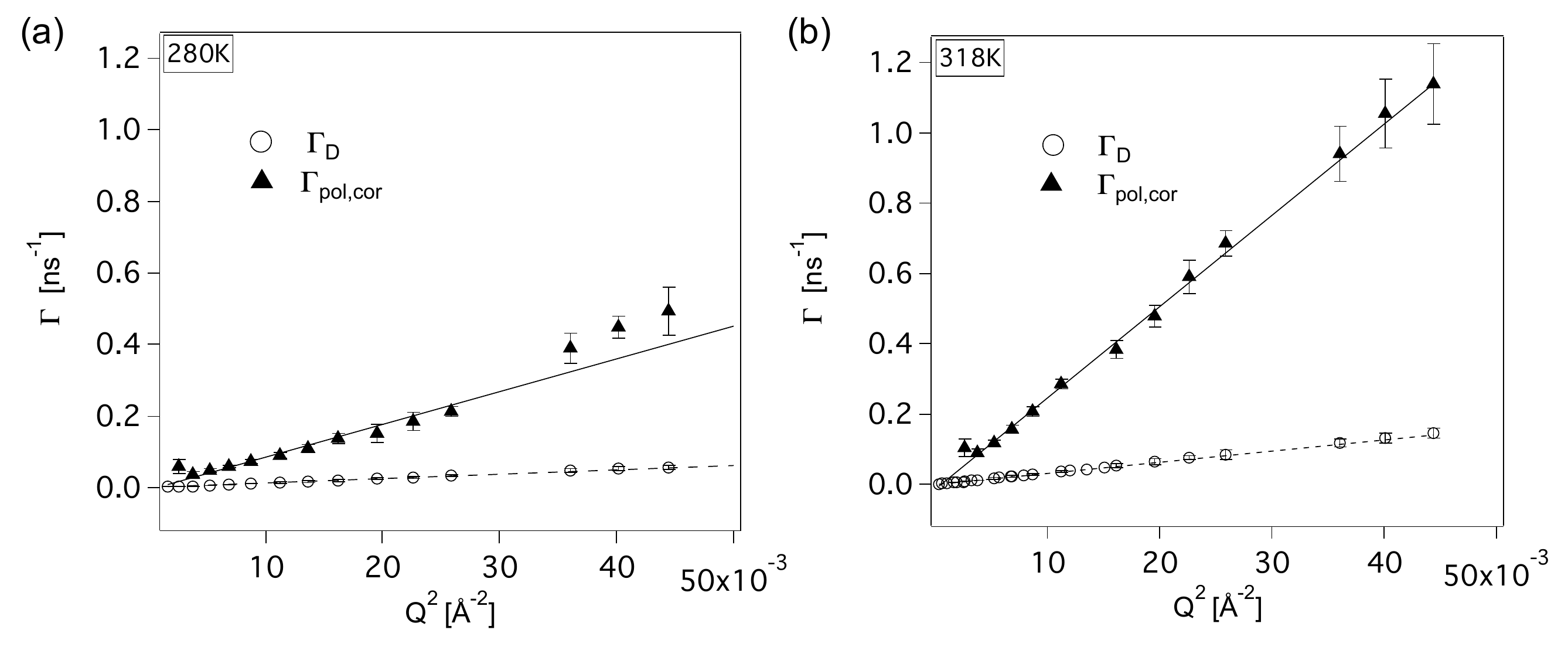}
 \caption{The values of  $\Gamma_D=1/\tau_D$ and $\Gamma_{pol,cor}$ as a function of $Q^2$ obtained for PEG2000 AuNP dispersed in D$_2$O  at the temperatures  T=280 K (a) T=318 K (b). }\label{fig:gammacorr}
\end{figure}

It is useful to plot the inverse of the relaxation times in terms of $\Gamma = 1/\tau$. As detailed in Eq. (\ref{IH_series}) the $\Gamma_{int} = \Bigg(DQ^2+\frac{1}{\tau_{\rm int,\it j}} \Bigg)$ contains also the term $DQ^2$ related to the translational diffusion of the whole NP. In Fig. \ref{fig:gammacorr} (a) and (b) we report as a function of $Q^2$ the $\Gamma_D=1/\tau_D$ and the $\Gamma_{pol,cor}$, where the latter corresponds to the internal dynamics corrected from  the translational diffusion of the whole NP. Both $\Gamma_D$ and $\Gamma_{pol,cor}$ have a clear linear trend that intercepts the origin. A dynamic whose relaxation can be described by a simple exponential decay in time, with a relaxation rate $\Gamma \sim Q^2$ has a diffusion-like behaviour. The coefficient diffusion associated to the relaxation time of the tethered polymer chains are around $9\pm2$ \AA$^2$ns$^{-1}$ and  $25\pm3$ \AA$^2$ns$^{-1}$ at temperatures of 280 and 318 K, respectively. Such a behaviour is also associated to a linear trend with $t$ of the mean square displacement. This  is in qualitative agreement with what observed in coarse grained molecular dynamics simulations \cite{LoVerso2013} that described the dynamics of brush polymers on spherical nanoparticles. This  is an important result since the fact that the dynamics of a corona is represented by a single exponential in this time and length-scale is not due to intrinsic limitation of the technique or of the analysis that are not able to distinguish a higher number of relaxations but it corroborates  computational studies. The linear trend for the mean square displacement found in reference \cite{LoVerso2013}  was qualitatively related to a Rouse description of a polymer brush under the boundary condition that constraints the tethered part of the polymer to the surface.\cite{johner1993} Under this assumption all the modes but one  are quickly damped. The only remaining mode has a relaxation behaviour that can effectively be described with a simple exponential. The diffusion constant associated to this dynamics was shown to be twice the diffusion constant of a free chain in solution.\cite{johner1993} This in fair agreement for example with the center of mass diffusion found for the PEG2000 solution of our study (see below).

If we tried to frame these dynamical results in a Zimm scenario the relaxation times of the polymers would be related to the viscosity, $\eta_s$, of the liquid medium by the following relation $\eta_{s,280K}/\eta_{s,318K}=(280\tau_1(Q)^{280K}/318\tau_1(Q)^{318K}) = R_z $.\cite{doi1988} The average ratio between the viscosity of the \dwat\ between the two temperatures 280 and 318 K is $\sim$ 2.6 whereas the $R_z\sim 2.3$.  Even if the trend of the relaxation times with temperature is roughly compatible with what the Zimm model predicts, the absolute value of the viscosity that it is associated with these relaxation times ($\eta_{solvent}=K_BT\tau_{pol}Q^3/6\pi$)\cite{kanaya2005} as predicted by the Zimm model is around 0.004 Pa$\cdot$s at 280 K and 0.0016 Pa$\cdot$s at 318K, i.e. roughly two times bigger than expected. A scaling of the relaxation time based only on the solvent viscosity neglects the important contribution to the dynamics related to the fact that the polymers are tethered to the NP surface. This has a significant impact on the resulting dynamics.  This fact suggests that our results cannot be fully represented in the context of this common used model.

This $\Gamma \sim Q^2$ behaviour  was already reported for amphiphilic diblock copolymer micelles in aqueous solution like in the work of  Matsuoka {\it et al.} \cite{matsuoka2000} and Castelletto {\it et al.}.\cite{castelletto2003}  In the latter, however, the $\Gamma$ corresponding to the internal dynamics of the micelle did not intercept the origin.  Our results are similar also to those found for polystirene-butadiene diblock copolymer micelles in that, the dynamical processes are ascribable to two relaxation times, one for the translational diffusion of the whole micelle and one accounting for the dynamics of the polymer corona. However, it differs in the fact that a Zimm behaviour was used to describe the latter. In agreement with all the above cited work the dynamics found for the polymer corona in this case is not compatible with the so called {\it breathing mode}, a collective mode of the polymer chains driven by a competition between the osmotic compressibility and the entropic force of the polymer chains predicted by de Gennes. \cite{degennes1986} Such a dynamics is characterised by a distribution of relaxation times, at variance with our experimental observation that gives evidence of only one relaxation time associated with the polymer chains. 

The fitting analysis of the NSE data for the PEG2000 solution dissolved in \dwat\ at 10\% in weight illustrates  also the presence of two dynamical processes with relaxation times $\tau_1$ and $\tau_2$ .  By a close inspection of Fig. \ref{fig:Figure-TauComparison}(a)  and \ref{fig:Figure-TauComparison}(b) we see that  the faster relaxation times $\tau_1$ measured for the homopolymer were very close to the relaxation times $\tau_{pol}$ obtained from the analysis of the NSE data of the PEG2000 AuNP at each $Q$. This is particularly evident at the highest temperature (Fig. \ref{fig:Figure-TauComparison}(b)). The slower dynamical process $\tau_2$  is compatible to the relaxation time associated to the translational diffusion of a PEG2000 polymer in \dwat\ as measured by NMR experiments at similar concentration.\cite{waggoner1995} A direct comparison is however not obvious since the data in literature are performed up to 8\% concentration and the extrapolation in these range of concentrations and temperatures are not trivial for the PEG2000. Therefore we tentatively ascribed this relaxation time to the translational dynamic of the polymer in the solution.  We will come back later to the consistency of this assumption.  	
	Previous attempts to fit the data corresponding to the PEG2000 homopolymers with a standard fitting procedure showed that these data could be fitted also using a single stretched exponential in the \textit{Q} range $0.06\div0.19$ \AA$^{-1}$. By letting the $\beta$-parameter free, the fit returned a value close to 0.9. A $\beta$ parameter smaller then one is a sign of a more complex dynamic scenario involving a distribution of relaxation times. \cite{johnston2006} Indeed, a distribution of relaxation times is very likely to occur in a macromolecular system such as a polymer in solution.   As we have seen previously in this section, instead of considering a continuous distribution of relaxation times, one can represent this complexity in a simplified way by describing the dynamical properties of the system with a finite number of components. But which is then the best way to describe our data?

With the aim of finding the model that support our data with the highest statistical significance, we then run the fitting analysis by letting  the RJ-MCMC algorithm use a model consisting of a linear combination of $\beta$-stretching exponentials (Eq. (\ref{eq:datamodel})) and leaving the algorithm free to estimate all the parameters. In looking for the best likelihood between the model and the data, the algorithm explored all the different possibilities such as sums of simple and/or stretched exponential functions, and clearly also  the single stretched exponential function that, as we have seen, can still describe the data. 

The prior distribution function for the $\beta_{j}$ (with $j=1\ldots,n$) described in Sec.\ref{sec:bayes} was used. Quite interestingly the algorithm converged to the same combination of simple exponentials described in section \ref{sec:PEG2000results}. The results found by the Bayesian algorithm showed quite clearly that the combination of one (only at very small $Q$) or two simple exponentials was indeed the most probable description of ${I}(Q,t)/{I}(Q,0)$. At each \textit{Q}, the algorithm reproduced within the errors the results  obtained for a combination of simple exponentials.  The estimated average values for $\beta_{1}$ and $\beta_{2}$ (when present) were both very close to one (typically $0.98/0.99\pm0.02$) and the posterior probability $P(\beta_{1}=1)$ and $P(\beta_{2}=1)$ were close to 100\% and between 85 and 96\%, respectively. 

More interestingly, the solution with two single exponential functions that picks out a finite number of relaxation processes was preferred over the one with one single stretched exponential that is  often used to describe the presence of multiple relaxation processes. We can conclude that the proposed method has allowed a sort of ``{\it{fine-structure}"} analysis of the dynamic scenario for the polymer solutions providing also a probabilistic statistical support to the model adopted for the decription of the time correlation  function relaxations. We stress that by choosing arbitrarily to fit the data with just one Kohlrausch-Williams-Watts (KWW)  function in aprioristic way,  part of the information contained in the data might be lost. Moreover in this particular case the choice of an apparent simpler model  (one single stretched exponential rather than two exponentials)  would not be even justified by some parameters economy criterion since in both cases the number of fitting parameters would be the same.   

This  analytical compact form is often used to represent NSE data in the contest of  the Zimm and the Rouse models,\cite{richter2005} where the Langevin equation is solved through an expansion in normal coordinates. A relaxation time is associated to each of these normal modes. The KWW results from an average over the normal modes used to solve the Langevin equation. Although a stretched exponential function is ruled out by the algorithm as highly improbable, this by no means implies that dynamic processes found in our analysis are at variance with what it is predicted by polymer theories. 

In fact, an interesting experimental result of our analysis of homopolymers solutions, is that the decay time ratio of the two exponential modes $\tau_1$ and $\tau_2$  found by the application of Eq. (\ref{eq:Intscatthomopol}) has a constant value of about 2 (to within experimental uncertainties), as displayed in Fig. 12.

\begin{figure}[htbp]
	\centering
	\includegraphics[width=110mm]{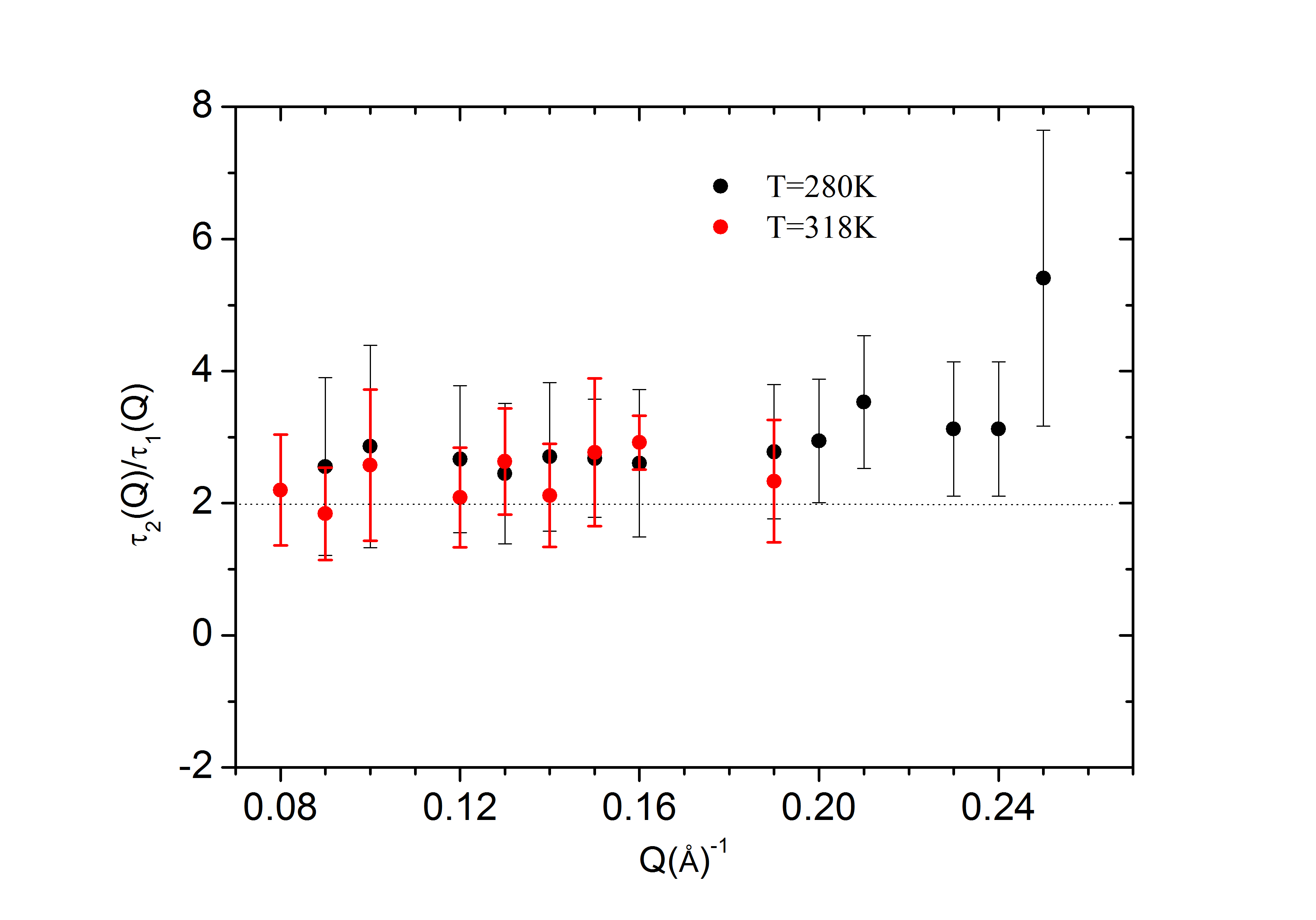}
	\caption{Ratio between the two relaxation times estimated by the RJ-MCMC algorithm for the PEG2000 solution, at the two investigated temperature. }	\label{fig:FigRousevsBayes}
\end{figure}

In the Rouse model \cite{doi1988,richter2005} the $p$=0 normal mode corresponds to the center of mass diffusion with a diffusion coefficient $D_R$. All other modes ($p\ge1$) account for the chain internal motions. Since the Bayesian analysis detected only two modes, including the overall polymer translational diffusion that we tentatively ascribed to the $\tau_2$ relaxation, we worked out the $I(Q,t)$\cite{richter2005} in the context of the Rouse model, taking into account only the first internal mode ($p=1$), that has the longest relaxation time.
In particular, it can be shown that $I'_{int}(Q,t)$, i.e. the time-dependent part  of the internal dynamics in  Eq.(\ref{IH_series}), which now reduces to a single exponential, contributes to $I(Q,t)/I(Q,0)$ with a term:

\begin{equation}\label{Rousemod}
exp(-D_R Q^2t)exp(-\frac{t}{\tau_{p=1}^{Rouse}})
\end{equation}

with

\begin{equation}
\tau_{p=1}^{Rouse}(Q)=\frac{3\pi^2}{Nb^2}\tau_R\frac{1}{Q^2}=\frac{1}{D_RQ^2}
\end{equation}

where $\tau_R=N^{2}b{^2}\zeta/3\pi^2k_{B}T$ is the Rouse time\cite{doi1988}, $N=45$ is the the number of monomer units of PEG2000, $b=3.5$ \AA \cite{Hansen2003} is the segment length in the Rouse chain and $\zeta$ is the microscopic friction coefficient.
Therefore we finally find that $I(Q,t)/I(Q,0)$ is given by the sum of two exponential terms with decay times expressed as $1/D_RQ^2$ and $1/2D_RQ^2$, respectively. Hence, in the context of Rouse Model and under the approximation used the ratio between the relaxation times $\tau_2$ and $\tau_1$ as expressed in Eq (\ref{eq:Intscatthomopol}) is 2, in fair agreement with the experimental finding.  Furthermore, this results provide us an internal consistency proof that $\tau_2$ can be confidently attributed to the polymer center of mass diffusion and allow us to calculate from their values an average values of $D_R \sim 12 \text{\AA}^2$/ns and $ 5\text{\AA}^2$/ns at 318 and 280K, respectively, and from the relation $\zeta=k_BT/ND_R$ a ratio $\zeta_{280K}/\zeta_{318K}\sim2.1$. If again a Rouse context is assumed to model the internal dynamics of the polymer chains in solution, the same ratio between the friction coefficients at the two investigated temperature can be calculated using the estimated relaxation times $\tau_{int,1}$ (see Eq. \ref{IH_series}) and they follow the temperature scaling relationship $280\cdot \tau_{int,1}(Q)^{280K}/318\cdot\tau_{int,1}(Q)^{318K} = \zeta_{280K}/\zeta_{318K} \sim 2.1$. This  shows a further internal consistency of the rescaling in temperatures of relaxation times according to the Rouse model.
	
Unlike for the PEG2000 AuNP, for the PEG 400 AuNP, ${I}(Q,t)/{I}(Q,0)$ could be always described only with one term corresponding to the translational dynamics of the NP, without the need of including a term accounting for the polymer dynamics. This could be due to different reasons. The net amount of the polymer, as well as the volume fraction of  polymer with respect to that of the whole NP is significantly lower for PEG400 AuNP compared to PEG2000 AuNP. As a matter of fact if we approximate the volume of the NP to $4/3\pi R_h^3$ and if we consider that the radius of the gold core is $R_{Au}= ~ 25$ \AA\, we have that in the case of the PEG2000 AuNP the hydrated polymer corona occupies $\sim$ 0.90 of the NP volume, whereas only 0.70 in the case of PEG400 AuNP. Moreover, the concentration of the NP in the solution were 5\% in weight for the PEG2000 AuNP and only 2\% in weight for the PEG400 AuNP. Clearly, the  contribution of the polymer corona in the two experiments was significantly different, in particular the scattering intensity due to the regions where  polymer was present was much higher in the case of the PEG2000 AuNP than in the case of the PEG400 AuNP. Furthermore, the condition $Q\cdot R_h >> 1$ that determines the \textit{Q} range  where the polymer dynamics dominates with respect to the translational diffusion is less effectively fulfilled in the case of the PEG400 AuNP due to the significantly smaller value of $R_h$. The combination of these two conditions might have obscured the scattering of the polymers in the case of the PEG400 AuNP. The lack of evidence of a polymer dynamics in our NSE experiment could also be due to an effect of confinement. The short PEG400 chains are anchored to the surface of the NP with an area per molecule around 14 \AA$^2$.\cite{maccarini2010} The effect of the anchor on a solid surface combined with the presence of neighbouring polymers highly packed  might slow down the fluctuation of the PEG400  to values outside the experimental window of the instrument.  Unlike the PEG400 that extends only up to 25 \AA\ away from the surface, the PEG2000  extends up to 60 \AA\ away from the surface allowing a significant part of the molecule to fluctuate in the larger volume available. 
  
 The NSE data measured on the 5\% solution of PEG400 in \dwat\  evidenced only one relaxation process in the time window of the experiment. This relaxation corresponds very well with the NMR results concerning the translational diffusion of the polymer at the same concentration and temperatures. \cite{waggoner1995}  A possible explanation for the absence of internal modes might be related to the fact that the relaxation time corresponding to the faster translational diffusion of PEG400 (compared to PEG2000) might be comparable with the relaxation time of the internal dynamics of PEG400. If the two processes occur at a very similar timescale it would be very difficult to distinguish between them.

\section{Conclusions}

We have exploited a Bayesian approach together with the very important property of decomposition of any time correlation function in terms of exponential functions to explore the nanosecond dynamics of polymer coated nanoparticles dissolved in \dwat\ in comparison with the dynamics of equivalent polymers free in solution measured through NSE spectroscopy. After decomposing the intermediate scattering function in elementary components, we associated each component to a particular kind of dynamics (center of mass translational diffusion or internal polymeric dynamics) both for the grafted nanoparticles and for the homopolymer solutions, by comparing the relaxation times obtained from the NSE analysis to those obtained with DLS and NMR data from literature and checking the agreement with polymer science theory. 
	For the PEGAuNP we disentangled the translational dynamics of the whole nanoparticle from that of the grafted polymer chains. The dynamic phenomena observed depended strongly on the molecular weight of the tethered polymer. In the case of longer PEG2000 grafted chains we could clearly see the translational dynamics of the NPs and the internal dynamics of the polymer chains. The polymers did not follow a Zimm dynamics in the range of times and momentum transfer explored, nor the often invoked breathing modes.  Instead our analysis indicates that the polymer corona relaxation follows a pure exponential decay, which appears to be an important confirmation of the behavior predicted by coarse grained molecular dynamics simulations \cite{LoVerso2013} and consistent with a Rouse description of a tethered polymer brush where all the Rouse modes are damped but the longest one related to  a movement over the entire polymer chain length. \cite{ johner1993} In the case of shorter PEG400 grafted chains, only the translational dynamics of the NP was visible. This could be due to the significantly lower amount of polymeric material (and the NSE signal associated with it) in this case, or due to a high mechanical constraint of the PEG400 in proximity of the gold surface.
	
	As for  polymer solutions, the analysis of the NSE data evidenced two relaxation times in the case of longer PEG2000 polymers associated to the translational dynamics of the polymer and  to the $p$=1 Rouse mode, respectively. The proposed analysis estabilishes that beside the center of mass molecule diffusion only one relaxation time is statistically supported and this is then necessarily described as a simple exponential decay. 
	As an important outcome the more common formal description of the time correlation function decay in terms of stretched exponential, as in the averaged form of Zimm and Rouse theory, is excluded by the joint posterior distribution of the adopted model conditional on the observed data. Nevertheless,  the internal relaxation time is absolutely consistent with the dominant Rouse time for a polymer free chain and effectively both estimated relaxations times scale with temperature according the Rouse theory.
	
	In the case of the shorter PEG400 only one relaxation time comparable with the translational diffusion of the polymer was evidenced. 
	
	Our methodology addresses one of the most important issues in the analysis of inelastic and quasi-elastic neutron scattering data - the arbitrariness in the choice of the number of components needed to describe the studied dynamical processes. In conventional fitting analysis, a particular model is imposed, based on a reasonable physical hypothesis and without any knowledge of a posterior probability that this choice of the model is statistically significant. Hence, the physical hypothesis, although well motivated, affects substantially the result of the fitting analysis. In the specific case of NSE experiments, nISF is often described in terms of stretched exponential that can account for the superposition of more relaxation mechanisms, in terms of an effective relaxation process modulated by a beta parameter. In view of our lack of knowledge about the most reliable model describing the data, no such invasive hypothesis was done to run the RJ-MCMC algorithm used in this work to explain the experimental data. We have shown for example how it is possible to make explicit the richness of the dynamic properties of a system such the PEG2000 solution that would be instead hidden if nISF's were modeled as a stretching exponential function. At the same time we provided a probabilistic evidence about the model used to describe the experimental data and without sacrificing the simplicity of the model description.    
	In doing so, we obtained that the output of the Bayesian analysis complies fully with the theoretically demonstrated general result on the exponential expansion of any time correlation function,\cite{bellissima,bellissima2,guarini_au}  which has been here applied for the first time to the description of the dynamics of a macromolecular system at the nanosecond time scale typically investigated by means of NSE.    
	We believe that the approach presented here will serve as an effective tool that can be used in general to provide an unbiased interpretation of neutron spin echo data.

\section{Acknowledgments}
The authors acknowledge the Institut Laue Langevin for neutron beam time allocated for the experiments. We acknowledge Forschungszentrum J\"ulich at the Heinz Maier- Leibnitz Zentrum (MLZ) in Garching for the allocation of beam time used to perform preliminary NSE experiments. In particular we thank Olaf Holderer and Michaela Zamponi for the assistance on the J-NSE instrument. We thank  Ralf Schweins for performing the DLS and SANS measurements. We thanks also Simone De Panfilis and Alessandro Paciaroni for very stimulating discussions. We are immensely grateful to Fabrizio Barocchi who provided insight and expertise that greatly strengthened important aspects of this work.
This article is dedicated to the memory of Giuseppe Briganti (1951-2015). 
\bibliographystyle{apsrev}
\bibliography{bibliography}

\pagebreak
\widetext
\begin{center}
\textbf{\large Supplemental Materials: A Bayesian analysis of neutron spin echo data on polymer coated gold nanoparticles in aqueous solutions}
\end{center}
\setcounter{equation}{0}
\setcounter{figure}{0}
\setcounter{table}{0}
\setcounter{page}{1}
\makeatletter
\renewcommand{\theequation}{S\arabic{equation}}
\renewcommand{\figurename}{Figure S}
\renewcommand{\tablename}{Table S}

\section{coherent and incoherent scattering}

\begin{figure}[htbp]
	\centering
	\includegraphics[width=90mm]{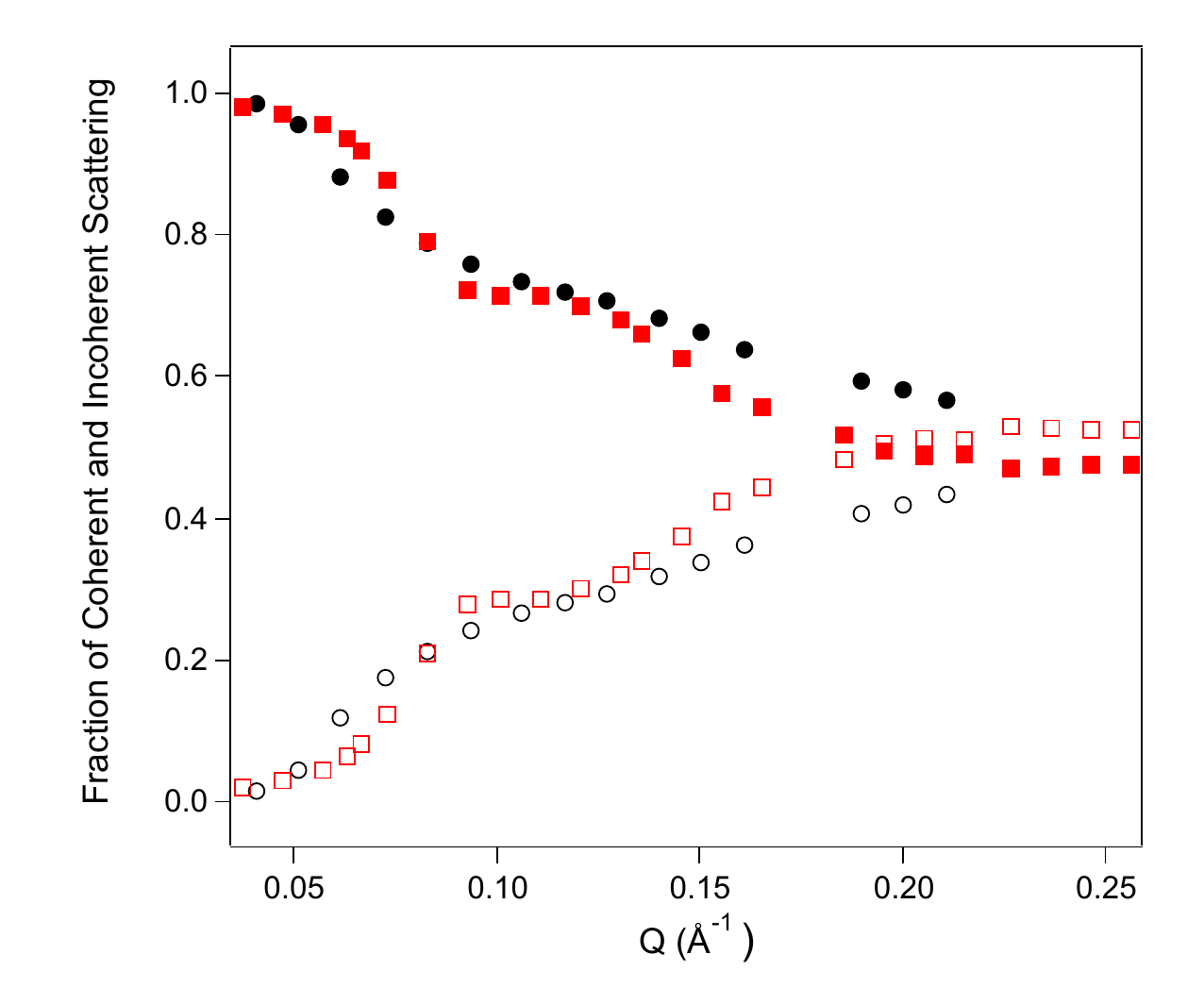}
	\caption{Fractions of coherent (filled symbols) and incoherent (empty symbols) contributions  to the NSE signal measured for PEG2000 AuNP (black circles) and PEG400 AuNP (red squares).}	\label{fig:Figure-Coh-Inc}
\end{figure}
\clearpage

\section{PEG2000 AuNP}
\begin{figure}[h]
\centering
\includegraphics[height=140mm]{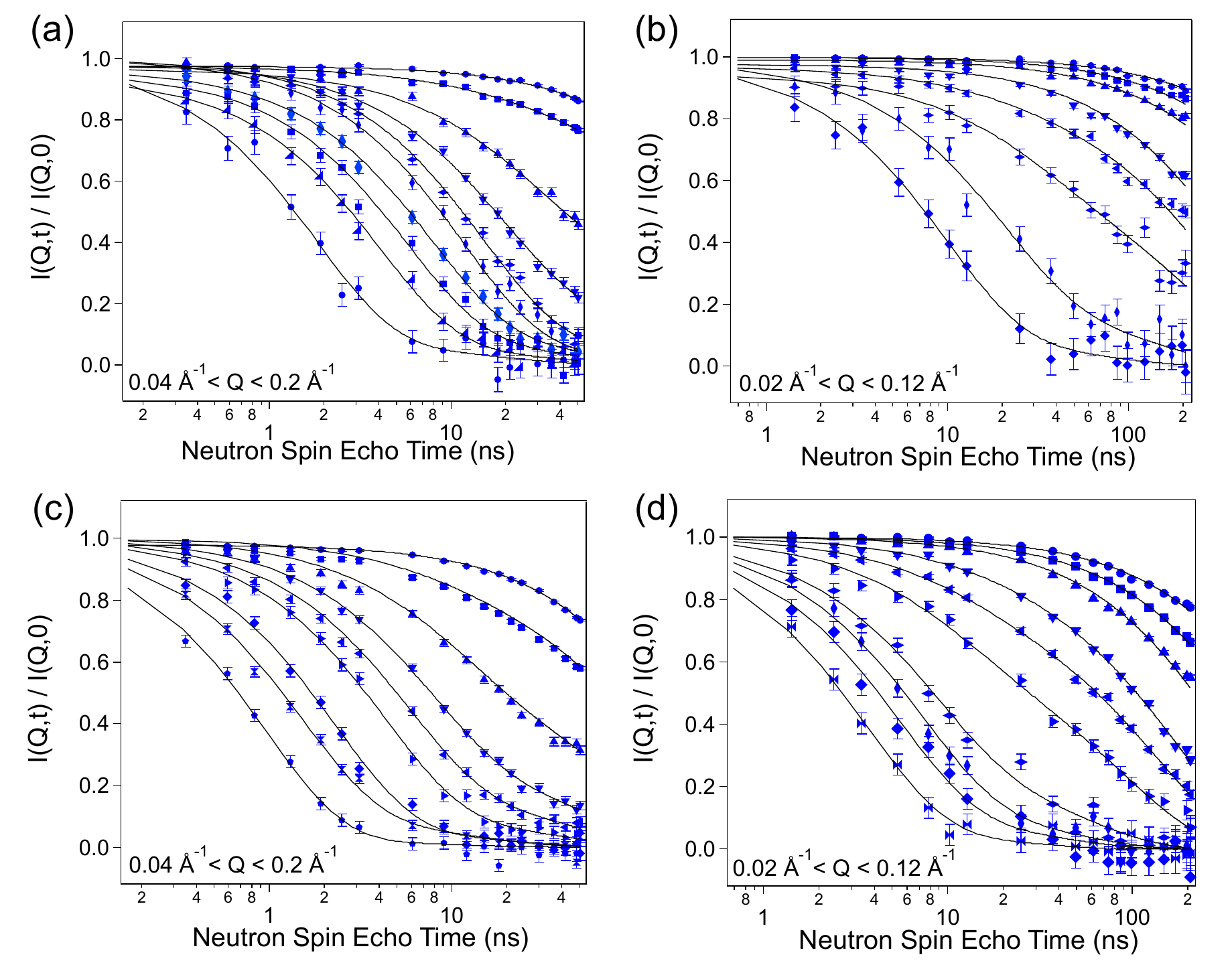}
 \caption{NSE curves representing I(Q,t)/I(Q,0) measured for PEG2000 AuNP dispersed in D$_2$O at different  temperatures and wavelengths. A) $\lambda = 10$ \AA, $T$ = 280 K;  B)  $\lambda = 16$ \AA, $T$ = 280 K; C) $\lambda = 10$ \AA, $T$ = 318 K D) $\lambda = 16$ \AA, $T$ = 318 K. The $Q$ values increase from above to bottom.}
\end{figure}
\begin{table}
	\centering
	\footnotesize
	\begin{tabular}{|c|c|c|c|c|c|c|c|} 
		\hline
		Q [\AA$^{-1}$]  &     $A_1$       &   $ \tau_D [\text{ns}] $  &  $ \tau_{pol} [\text{ns}]$   &   k   & P($k | y$) &  $D_T [\text{\AA}^2\text{ns}^{-1}]$   \\ \hline \hline
		0.04  &  1  &  $430\pm10$  &     &  1  &  0.82  &  $1.42\pm0.06$  \\ \hline
		0.05  &  $0.93\pm0.02$  &  $300\pm10$  &  $16\pm5$  &  2  &  1  &  $1.28\pm0.07$  \\ \hline
		0.06  & $0.54\pm0.05$  &  $200\pm20$  & $24\pm5$  &  2  &  1  &  $1.3\pm0.1$  \\ \hline
		0.07  & $0.25\pm0.04$  &  $150\pm10$  &  $18\pm2$  &  2  &  0.51  &  $1.3\pm0.1$  \\ \hline
		0.08  &  $0.10\pm0.03$  &  $110\pm10$  &  $15\pm0.7$  & 2  &  1  &  $1.3\pm0.1$  \\ \hline
		0.09  &  $0.07\pm0.02$  &  $89\pm6$  & $11.7\pm0.6$  & 2  &  0.56  &  $1.3\pm0.1$  \\ \hline
		0.11  &  $0.05\pm0.02$  &  $69\pm7$   & $9.5\pm0.5$  &   2  &  0.69  &  $1.3\pm0.1$  \\ \hline
		0.12  &  $0.10\pm0.3$  &  $56\pm6$  & $7.8\pm0.5$  &  2  &  0.72  &  $1.3\pm0.1$  \\ \hline
		0.13  &  $0.11\pm0.3$ &  $48\pm5$  &  $6.3\pm0.5$  &  2  &  0.73  &  $1.3\pm0.1$  \\ \hline
		0.14  &  $0.10\pm0.04$  &  $39\pm4$  &  $5.6\pm0.7$  & 2  &  0.72  &  $1.3\pm0.1$  \\ \hline
		0.15  & $0.09\pm0.03$  &  $34\pm3$  &  $4.6\pm0.5$  &  2  &  0.86  &  $1.3\pm0.1$  \\ \hline
		0.16  & $0.08\pm0.02$  &  $29\pm2$  &  $4\pm0.2$  &  2  &  0.83  &  $1.3\pm0.1$  \\ \hline
		0.19  & $0.06\pm0.02$  &  $21\pm2$  &  $2.3\pm0.2$  &   2  &  0.93  &  $1.3\pm0.1$  \\ \hline
		0.20  &  $0.06\pm0.02$  &  $19\pm2$  & $2.0\pm0.1$  &  2  &  0.93  &  $1.3\pm0.1$  \\ \hline
		0.21  &  $0.07\pm0.04$  &  $17\pm2$  &  $1.8\pm0.2$  &  2  &  0.94  &  $1.3\pm0.1$  \\ \hline
	\end{tabular}
	\caption{Fitting parameters obtained for the NSE measurements taken on the PEG2000 AuNP at  280 K and at $\lambda= 10$ \AA. P($k | y$) is the probability for the most frequently chosen model. The constant $\gamma$ (that was not reported in this table and in the next ones) in front of Eq. 10 in the manuscript, resulted always in the range 0.98--1 with few exceptions within a 5\% difference from 1.}
	\label{tab:PEG2000AuNP280K10A} 
\end{table}
\begin{table}
\centering
\footnotesize
\begin{tabular}{|c|c|c|c|c|c|c|} 
\hline
	Q [\AA$^{-1}$]  &     $A_1$       &   $ \tau_D [\text{ns}] $  &  $ \tau_{pol} [\text{ns}]$   &   k   & P($k | y$) &  $D_T [\text{\AA}^2\text{ns}^{-1}]$   \\ \hline \hline

  0.019  &  1      &  $1890\pm50$  &          &  1  &  0.98  &  $1.41\pm0.04$  \\ \hline
  0.026  &  1      &  $1250\pm40$  &          &  1  &  0.97  &  $1.18\pm0.04$  \\ \hline
  0.032  &  1      &  $860\pm40$   &          &  1  &  0.87  &  $1.11\pm0.05$  \\ \hline
  0.044  &  1      &  $400\pm10$   &          &  1  &  0.79  &  $1.29\pm0.04$  \\ \hline
  0.050  &  $0.90\pm0.02$  &  $310\pm7$    &  $20\pm7$&  2  &  0.91  &  $1.30\pm0.03$  \\ \hline
  0.057  &  $0.66\pm0.04$  &  $230\pm10$   &  $28\pm9$&  2  &  0.59  &  $1.31\pm0.07$  \\ \hline
  0.076  &  $0.22\pm0.06$  &  $132\pm7$    &  $20\pm4$&  2  &  0.55  &  $1.32\pm0.07$  \\ \hline
  0.082  &  $0.09\pm0.04$  &  $111\pm6$    &  $16\pm2$&  2  &  0.74  &  $1.32\pm0.07$  \\ \hline
  0.089  &  $0.19\pm0.05$  &  $96\pm5$     &  $13\pm2$&  2  &  0.68  &  $1.32\pm0.07$  \\ \hline
  0.110  &  $0.10\pm0.04$  &  $62\pm3$     &  $9.7\pm0.9$  &  2  &  0.74  &  $1.32\pm0.07$  \\ \hline
  0.116  &  $0.13\pm0.05$  &  $56\pm3$     &  $9\pm1$  &  2  &  0.78  &  $1.32\pm0.07$ \\ \hline
  0.123  &  $0.22\pm0.07$  &  $50\pm3$     &  $8\pm2$  &  2  &  0.70  &  $1.32\pm0.07$  \\ \hline
  \end{tabular}
 \caption{Fitting parameters obtained for the NSE measurements taken on the PEG2000 AuNP at  280K and at $\lambda= 16$ \AA.} 
   \label{tab:PEG2000AuNP280K16A} 
\end{table}
\begin{table}
	\centering
	\footnotesize
	\begin{tabular}{|c|c|c|c|c|c|c|c|c|} 
		\hline
		Q [\AA$^{-1}$]  &    $A_1$       &   $ \tau_D [\text{ns}] $  & $ \tau_{pol} [\text{ns}]$  & $\beta_2$    & P($\beta_2=1 |  k=2$)   & k   & P($k | y$) &  $D_T [\text{\AA}^2\text{ns}^{-1}]$  \\ \hline \hline
		0.041  &  1	  &  $177\pm2$ &          &     &              &  1  &  0.80  &  $3.42\pm0.09$\\ \hline
		0.051  &  $0.86\pm0.02$&  $128\pm9$ &  $9\pm2$ &  1  & 1*  &  2  &  1  & $3.0\pm0.2$\\ \hline
		0.062  &  $0.55\pm0.03$&  $87\pm9$  &  $10\pm1$&  1  & 1*   &  2 &  1      &  $3.1\pm0.3$  \\ \hline
		0.072  &  $0.28\pm0.02$  & $60\pm6$  & $7.5\pm0.3$   &  $1.000\pm0.005$  &  0.97  &  2  &  0.77  & $3.2\pm0.3$  \\ \hline
		0.083  &  $0.18\pm0.02$  & $47\pm5$  & $5.6\pm0.3$  &  1  & 1*  &  2  &  1  &  $3.2\pm0.3$  \\ \hline
		0.094  &  $0.11\pm0.02$  & $36\pm4$  & $4.2\pm0.2$  &  $1.000\pm0.006$  &  0.97  &  2  &  0.84  &  $3.2\pm0.3$  \\ \hline
		0.106  &  $0.04\pm0.01$  & $27\pm3$  & $3.1\pm0.1$  &  $0.99\pm0.02$  &  0.83  &  2  &  0.98  &  $3.3\pm0.3$  \\ \hline
		0.127  &  $0.06\pm0.02$  & $19\pm2$  & $2.3\pm0.1$  & $0.999\pm0.008$  &  0.96  &  2  &  0.98  &  $3.3\pm0.3$  \\ \hline
		0.140  &  $0.08\pm0.02$  & $15\pm2$  & $1.8\pm0.1$  & $0.99\pm0.03$  &  0.86  &  2  &  0.98  &  $3.4\pm0.3$  \\ \hline
		0.150  &  $0.10\pm0.03$  & $13\pm1$  & $1.50\pm0.09$  & $0.99\pm0.03$ &  0.80  &  2  &  0.98  &  $3.4\pm0.3$  \\ \hline
		0.161  &  $0.07\pm0.02$  & $12\pm2$  & $1.30\pm0.08$  & $0.99\pm0.02$  &  0.91  &  2  &  0.98  &  $3.3\pm0.3$  \\ \hline
		0.190  &  $0.03\pm0.02$  & $8.5\pm0.9$ & $0.95\pm0.05$  &$0.99\pm0.02$ &  0.89  &  2  &  0.99  &  $3.3\pm0.3$  \\ \hline
		0.200  &  $0.04\pm0.02$  & $7.6\pm0.8$  &$0.84\pm0.06$  &$0.99\pm0.02$ &  0.91  &  2  &  0.99  &  $3.3\pm0.3$  \\ \hline
		0.211  &  $0.03\pm0.02$  & $6.9\pm0.7$  &$0.78\pm0.06$  &$0.99\pm0.02$ &  0.91  &  2  &  0.99  &  $3.3\pm0.3$  \\ \hline 
	\end{tabular}
	\caption{Fitting parameters obtained for the NSE measurements taken on the PEG2000 AuNP  at  318 K and at $\lambda$= 10 \AA. By way of example in the 5$^{th}$ and the 6$^{th}$ column of this table we report also the average estimated value of $\beta_2$ and the probability to have $\beta_2=1$ when the two component model is chosen. \\{\footnotesize  $^{*}$ For these datasets we set $k=2$, for consistency with the results obtained for the other data sets, even if the model with two components is not the one with the highest posterior. Moreover the joint posterior distribution (conditional on $k=2$) is bimodal. The first maximum in the distribution corresponds to $\beta_2=1$ and values of the other parameters as given in the table. The second maximum corresponds to values of the parameters that are meaningless from a physical point of view. For example, at $Q$ =0.051 \AA$^{-1}$, the second mode corresponds to $\beta_2=0.71$, $A_1=0.27$, $\tau_D$ = 131.12 ns and $\tau_{pol}$= 86.55 ns, i.e. values of the parameters completely incongruous with the trend observed for the other datasets. This second mode is probably a spurious solution, related to a local maximum and was, therefore, discarded. Thus, we simply set $\beta_2=1$ for these datasets, in analogy with the values estimated for the remaining datasets. In all the other datasets we reported in the table the average value of $\beta_{2}$ when $k=2$ as estimated by the algorithm but a value of 1 for the stretching coefficient when the conditional probability  $P(\beta_2=1 |  k=2)>0.5$ was adopted for the final curve fitting.}}
	\label{tab:PEG2000AuNP316K10A} 
\end{table}
\begin{table}
\centering
\footnotesize
\begin{tabular}{|c|c|c|c|c|c|c|c|c|} 
\hline
     Q [\AA$^{-1}$]  &    $A_1$       &   $ \tau_D [\text{ns}] $  & $ \tau_{pol} [\text{ns}]$    & k   & P($k | y$) &  $D_T [\text{\AA}^2\text{ns}^{-1}]$  \\ \hline \hline

  0.019  &   1  &  $760\pm10$  &     &                       1  &  0.76  &  $3.5\pm0.2$  \\ \hline
  0.026  &   1  &  $488\pm4$   &     &                       1  &  0.58  &  $3.2\pm0.3$  \\ \hline
  0.032  &   1  &  $312\pm4$   &     &                       1  &  0.75  &  $3.0\pm0.1$  \\ \hline
  0.044  &   $0.92\pm0.02$  &  $175\pm6$   & $16\pm5$    &   2  &  1*     &  $3.1\pm0.3$  \\ \hline
  0.051  &   $0.80\pm0.02$  &  $135\pm5$  &  $12\pm2$    &   2  &  1*     &  $3.1\pm0.3$  \\ \hline
  0.057  &   $0.64\pm0.04$  &  $92\pm7$   &  $10\pm2$    &   2  &  1*    &  $3.4\pm0.3$  \\ \hline
  0.076  &   $0.28\pm0.06$  &  $52\pm5$   &  $8 \pm1$    &   2  &  0.92  &  $3.4\pm0.3$  \\ \hline
  0.082  &   $0.14\pm0.05$  &  $46\pm5$   &  $7.1\pm0.6$ &   2  &  0.93  &  $3.3\pm0.3$  \\ \hline
  0.089  &   $0.10\pm0.04$  &  $38\pm4$  &  $5.3\pm0.6$  &   2  &  0.97  &  $3.4\pm0.3$  \\ \hline
  0.110  &   $0.06\pm0.03$  &  $26\pm3$  &  $3.6\pm0.2$  &   2  &  0.94  &  $3.2\pm0.2$  \\ \hline
  0.116  &   $0.11\pm0.04$  &  $23\pm2$  &  $3.1\pm0.3$  &   2  &  0.97  &  $3.3\pm0.3$  \\ \hline
  0.123  &   $0.10\pm0.04$  &  $21\pm2$  &  $2.7\pm0.3$  &   2  &  0.85  &  $3.2\pm0.3$  \\ \hline  \end{tabular}
 \caption{Fitting parameters obtained for the NSE measurements taken on the PEG2000 AuNP  at  318 K and at $\lambda= 16$ \AA. *Refer to note in Table \ref{tab:PEG2000AuNP316K10A}.}
   \label{tab:PEG2000AuNP316K16A} 
\end{table}

\clearpage

\begin{figure}[htbp]
	\centering
	\includegraphics[width=140mm]{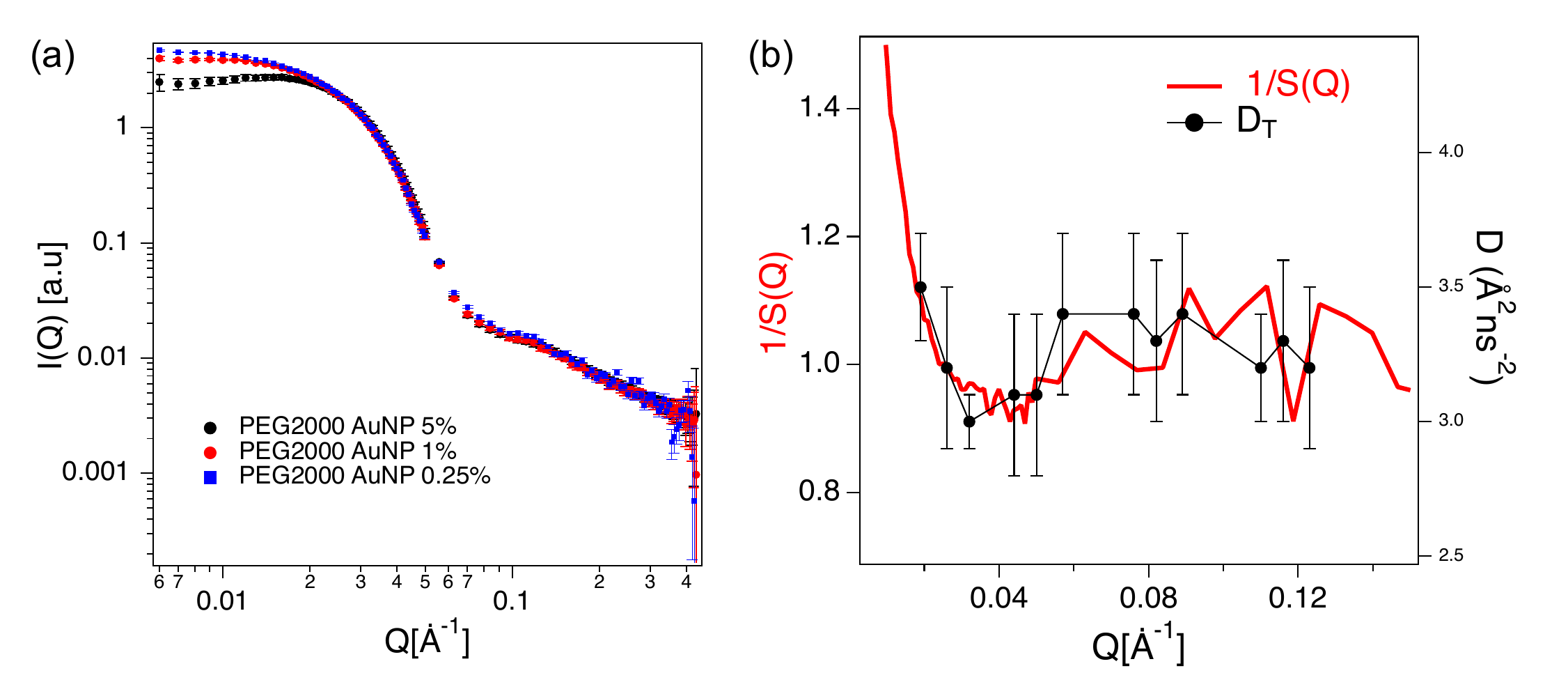}
	\caption{(a) SANS curves normalised to the concentration of the PEG2000 Au NP. (b) Inverse of the structure factor (1/S(Q)) extracted from the SANS curves at different concentrations and the translational diffusion D obtained by the Bayesian analysis.}	
\end{figure}
\clearpage

\clearpage

\section{PEG400 AuNP}
\begin{figure}[h]
\centering
\includegraphics[height=120mm]{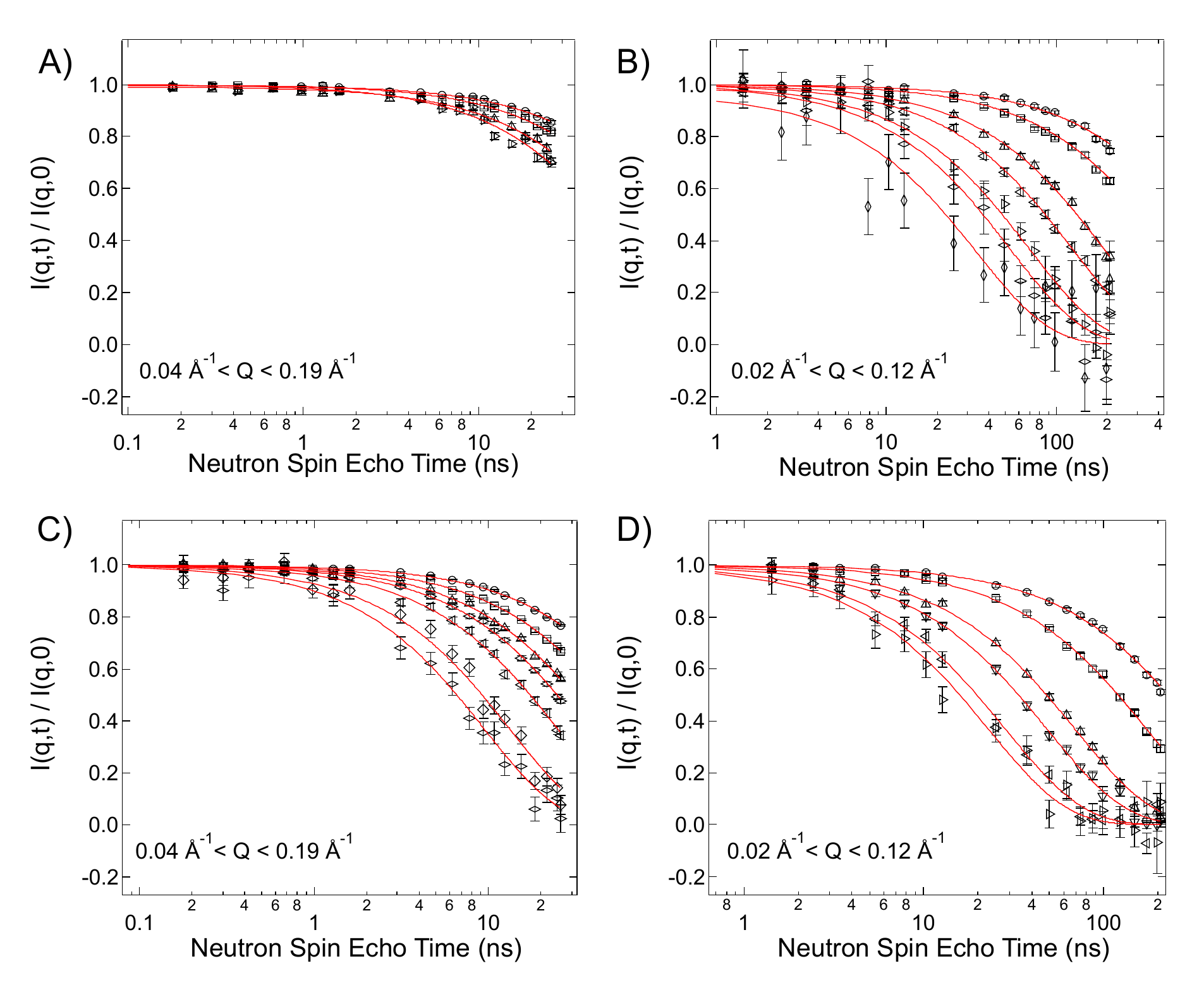}
 \caption{NSE curves representing $I(Q,t)/I(Q,0)$ measured on the PEG400 AuNP at different temperatures and wavelengths. A) $\lambda$ = 8 \AA, $T$ = 280 K;  B)  $\lambda$ = 16 \AA, $T$ = 280 K; C) $\lambda$ = 8 \AA, $T$ = 318 K D) $\lambda$ = 16 \AA, $T$ = 318 K. The $Q$ values increase from above to bottom.}\label{fig:PEG400AuNP}
\end{figure}


\begin{table}[h]
\centering
\footnotesize
\begin{tabular}{|c|c|c|c|c|c|} 
\hline
   Q [\AA$^{-1}$]  &     $A_1$       &   $ \tau_D [\text{ns}] $  &  k   & P($k | y$) &  $D_T [\text{\AA}^2\text{ns}^{-1}]$ \\ \hline \hline
 
  0.047  &   1  &  $173\pm7$ &   1  &  0.69  &  $2.6\pm0.2$  \\ \hline
  0.057  &   1  &  $128\pm5$ &   1  &  0.35$^*$  &  $2.4\pm0.2$  \\ \hline
  0.067  &   1  &  $90\pm3$  &   1  &  0.57  &  $2.4\pm0.2$  \\ \hline
  0.063  &   1  &  $93\pm3$  &   1  &  0.83  &  $2.6\pm0.3$  \\ \hline
  0.073  &   1  &  $71\pm2$  &   1  &  0.84  &  $2.6\pm0.2$  \\ \hline
  0.083  &   1  &  $58\pm5$  &   1  &  0.59  &  $2.6\pm0.3$  \\ \hline
  0.093  &   1  &  $42\pm5$  &   1  &  0.79  &  $2.4\pm0.3$  \\ \hline  
        \end{tabular}
 \caption{Fitting parameters obtained for the NSE measurements on the PEG400 AuNP taken at  280 K and at $\lambda$= 8 \AA. \\
 $^*$ For this particular dataset  the data are scarcely informative about the number of components and the algorithm assigns relevant posterior probabilities to models with either 1, 2 or 3 components. Therefore, in this case, model choice takes advantage from results obtained for similar data sets, for which the posterior distribution of the number of components is more peaked and the model with 1 component is largely supported.}
   \label{tab:PEG400AuNP280K8A} 
\end{table}

\begin{table}[h]
\centering
\footnotesize
\begin{tabular}{|c|c|c|c|c|c|} 
\hline
   Q [\AA$^{-1}$]  &    $A_1$       &   $ \tau_D [\text{ns}] $  &  k   & P($k | y$) &  $D_T [\text{\AA}^2\text{ns}^{-1}]$ \\ \hline \hline  0.02  &  1  &  $800\pm20$ &   1  &  0.89  & $3.1\pm0.1$  \\ \hline
  0.03  &  1  &  $458\pm9$  &   1  &  0.59  & $2.5\pm0.1$  \\ \hline
  0.05  &  1  &  $191\pm3$  &   1  &  0.78  & $2.5\pm0.1$  \\ \hline
  0.06  &  1  &  $124\pm2$  &   1  &  0.71  & $2.4\pm0.3$  \\ \hline
  0.07  &  1  &  $69\pm3$   &   1  &  0.91  & $2.7\pm0.3$  \\ \hline
  0.08  &  1  &  $54\pm4$   &   1  &  0.92  & $2.7\pm0.2$  \\ \hline
  0.11  &  1  &  $34\pm5$   &   1  &  0.69  &  $2.4\pm0.4$   \\ \hline
   
    \end{tabular}
 \caption{Fitting parameters obtained for the NSE measurements on the PEG400 AuNP taken at  280 K and at $\lambda$ = 16 \AA. }
   \label{tab:PEG400AuNP280K16A} 
\end{table}

\begin{table}[h]
\centering
\footnotesize
\begin{tabular}{|c|c|c|c|c|c|} 
\hline
   Q [\AA$^{-1}$]  &     $A_1$      &   $ \tau_D [\text{ns}] $  &  k   & P($k | y$) &  $D_T [\text{\AA}^2\text{ns}^{-1}]$ \\ \hline \hline  0.02  &  1  &  $800\pm20$ &   1  &  0.89  & $3.1\pm0.1$  \\ \hline
  0.037  & 1  &  $99\pm1$      & 1  &  0.87   &  $7.2\pm0.4$  \\ \hline
  0.048  & 1  &  $64.4\pm0.8$  & 1  &  0.89   &  $6.8\pm0.4$  \\ \hline
  0.057  & 1  &  $44.7\pm0.6$  & 1  &  0.91   &  $6.8\pm0.5$  \\ \hline
  0.067  & 1  &  $31.6\pm0.5$  & 1  &  0.95  &  $7.1\pm0.2$  \\ \hline
  0.063  & 1  &  $35.7\pm0.4$  & 1  &  0.94  &  $7.0\pm0.1$  \\ \hline
  0.073  & 1  &  $25.0\pm0.4$  & 1  &  0.96  &  $7.5\pm0.3$  \\ \hline
  0.083  & 1  &  $19.3\pm0.5$  & 1  &  0.91  &  $7.6\pm0.6$  \\ \hline
  0.093  & 1  &  $17.1\pm0.7$  & 1  &  0.90  &  $6.9\pm0.4$  \\ \hline
  0.100  & 1  &  $15.0\pm0.5$  & 1  &  0.92  &  $6.6\pm0.3$  \\ \hline
  0.111  & 1  &  $13.2\pm0.5$  & 1  &  0.96  &  $6.2\pm0.3$  \\ \hline
  0.121  & 1  &  $10.8\pm0.4$  & 1  &  0.97  &  $6.4\pm0.3$  \\ \hline
  0.131  & 1  &  $9.5\pm0.4$   & 1  &  0.96  &  $6.2\pm0.3$  \\ \hline
  0.136  & 1  &  $8.7\pm0.4$   & 1  &  0.94  &  $6.2\pm0.5$  \\ \hline
  0.146  & 1  &  $7.5\pm0.4$   & 1  &  0.96  &  $6.3\pm0.3$  \\ \hline
  0.155  & 1  &  $8.3\pm0.9$   & 1  &  0.83  &  $5.3\pm0.6$  \\ \hline
  0.165  & 1  &  $6.6\pm0.9$   & 1  &  0.92  &  $5.6\pm0.7$  \\ \hline
  0.186  & 1  &  $4.7\pm0.8$   & 1  &  0.93  &  $6\pm1$  \\ \hline
        \end{tabular}
 \caption{Fitting parameters obtained for the NSE measurements on the PEG400 AuNP taken at  318 K and at $\lambda$ = 8 \AA.}
   \label{tab:PEG400AuNP318K8A} 
\end{table}

\begin{table}[h]
\centering
\footnotesize
\begin{tabular}{|c|c|c|c|c|c|} 
\hline
   Q [\AA$^{-1}$]  &     $A_1$       &   $ \tau_D [\text{ns}] $  &  k   & P($k | y$) &  $D_T [\text{\AA}^2\text{ns}^{-1}]$ \\ \hline \hline  0.02  &  1  &  $800\pm20$ &   1  &  0.89  & $3.1\pm0.1$  \\ \hline
  0.02  &  1  &  $329\pm4$  &  1  &  0.74  &  $7.6\pm0.2$    \\ \hline
  0.03  &  1  &  $173\pm2$  &  1  &  0.87  &  $6.6\pm0.4$    \\ \hline
  0.05  &  1  &  $69\pm1$   &  1  &  0.94  &  $6.8\pm0.3$    \\ \hline
  0.06  &  1  &  $48\pm1$   &  1  &  0.93  &  $6.6\pm0.2$    \\ \hline
  0.07  &  1  &  $28\pm1$   &  1  &  0.94  &  $6.7\pm0.3$    \\ \hline
  0.08  &  1  &  $23\pm2$   &  1  &  0.85  &  $6.5\pm0.6$    \\ \hline
  0.11  &  1  &  $12\pm2$   &  1  &  0.86  &  $6.8\pm0.8$    \\ \hline
  0.12  &  1  &  $10\pm2$   &  1  &  0.87   &  $6.7\pm0.9$    \\ \hline  
\end{tabular}
 \caption{Fitting parameters obtained for the NSE measurements on the PEG400 AuNP taken at  318 K and at $\lambda$ = 16 \AA.}
   \label{tab:PEG400AuNP318K16A} 
\end{table}

\clearpage

\section{PEG2000}
\begin{table}[h]
\centering
\footnotesize
\begin{tabular}{|c|c|c|c|c|c|c|c|c|c|} 
\hline
   Q [\AA$^{-1}$]  &     $A_1$       &   $ \tau_1 [\text{ns}] $   &   $ \tau_2$ [\text{ns}]   & k   & P($k | y$)   \\ \hline \hline
  0.06  &  1   &  24.3  $\pm$  0.4       &        &            1  &  0.75  \\ \hline
  0.08  &  1   &  17.7  $\pm$  0.2       &        &            1  &  0.65  \\ \hline
  0.09  & $0.5\pm0.2$ &  9  $\pm$  2     &  23  $\pm$  7    &  2  &  0.64  \\ \hline
  0.1   & $0.6\pm0.2$ &  7  $\pm$  2     &  20  $\pm$  5    &  2  &  0.64  \\ \hline
  0.12  & $0.6\pm0.2$  &  6  $\pm$  1    &  16  $\pm$  4    &  2  &  0.65  \\ \hline
  0.13  & $0.6\pm0.2$ &  4.9  $\pm$  0.9 &  12  $\pm$  3    &  2  &  0.63  \\ \hline
  0.14  & $0.5\pm0.2$ &  3.7  $\pm$  0.8 &  10  $\pm$  2    &  2  &  0.61  \\ \hline
  0.15  & $0.4\pm0.1$ &  2.8  $\pm$  0.6 &  7.5  $\pm$  0.9 &  2  &  0.62  \\ \hline
  0.16  & $0.4\pm0.2$  &  2.3  $\pm$  0.6&  6  $\pm$  1     &  2  &  0.57  \\ \hline
  0.19  & $0.5\pm0.2$ &  1.8  $\pm$  0.3 &  5  $\pm$  1     &  2  &  0.63  \\ \hline
  0.2   & $0.5\pm0.1$ &  1.8  $\pm$  0.3 &  5.3  $\pm$  0.8 &  2  &  0.61  \\ \hline
  0.21  & $0.6\pm0.1$ &  1.7  $\pm$  0.2 &  6  $\pm$  1     &  2  &  0.61  \\ \hline
  0.23  & $0.69\pm0.07$ &  1.6  $\pm$  0.2& 5  $\pm$  1     &  2  &  0.66  \\ \hline
  0.24  & $0.7\pm0.1$  &  1.6  $\pm$  0.2 & 5  $\pm$  1     &  2  &  0.67  \\ \hline
  0.25  & $0.69\pm0.05$ &  1.11  $\pm$  0.09&  6  $\pm$  2  &  2  &  0.73  \\ \hline
 
\end{tabular}
 \caption{Fitting parameters obtained for the NSE measurements on the PEG2000 in \dwat\ at concentration of 10\% in weight  taken at  280 K and at $\lambda$ = 8 \AA.}
   \label{tab:PEG2000280K8A10pc} 
\end{table}

\begin{table}[h]
\centering
\footnotesize
\begin{tabular}{|c|c|c|c|c|c|c|c|c|c|} 
\hline
     Q [\AA$^{-1}$]  &       $A_1$       &   $ \tau_1 [\text{ns}] $  & $ \tau_{2}[\text{ns}]$   &  k   & P($k | y$)   \\ \hline \hline
  0.06  &   1           &  10.7 $\pm$  0.1   &               &  1  &  0.55  \\ \hline
  0.08  &  $0.5\pm0.2$  &  5    $\pm$  1     &   11$\pm$  2  &  2  &  0.64  \\ \hline
  0.09  &  $0.5\pm0.2$  &  3.8  $\pm$  0.9   &   7$\pm$  1   &  2  &  0.63  \\ \hline
  0.1   &   $0.7\pm0.1$ &  3.3  $\pm$  0.3   &   8.5$\pm$3   &  2  &  0.68  \\ \hline
  0.12  &  $0.6\pm0.2$  &  2.3  $\pm$  0.4   &   4.8$\pm$0.9 &  2  &  0.72  \\ \hline
  0.13  &  $0.7\pm0.1$  &  1.9  $\pm$  0.2   &   5  $\pm$  1 &  2  &  0.73  \\ \hline
  0.14  &  $0.7\pm0.2$  &  1.7  $\pm$  0.2   &   3.6$\pm$0.9 &  2  &  0.78  \\ \hline
  0.15  &  $0.7\pm0.1$  &  1.3  $\pm$  0.2   &   3.6$\pm$0.9 &  2  &  0.81  \\ \hline
  0.16  &  $0.8\pm0.1$  &  1.2  $\pm$  0.1   &   3.5$\pm$0.2 &  2  &  0.81  \\ \hline
  0.19  &  $0.7\pm0.2$  &  0.9  $\pm$  0.1   &   2.1$\pm$0.6 &  2  &  0.75  \\ \hline
  0.2   &  $0.90\pm0.05$& 0.89  $\pm$  0.02  &   -           &  2  &  0.82  \\ \hline
  0.21  &  $0.97\pm0.03$& 0.73  $\pm$  0.04  &   -           &  1  &  0.53  \\ \hline
  0.23  &  $0.93\pm0.05$& 0.64  $\pm$  0.02  &   -           &  2  &  0.85  \\ \hline
  0.24  &  $0.95\pm0.05$& 0.52  $\pm$  0.02  &   -           &  2  &  0.91  \\ \hline
  0.25  &  $0.94\pm0.05$& 0.44  $\pm$  0.02  &   -           &  2  &  0.93  \\ \hline   \end{tabular}
 \caption{Fitting parameters obtained for the NSE measurements on the PEG2000 in \dwat\ at concentration of 10\% in weight taken at 318 K and at $\lambda$ = 8 \AA.}
   \label{tab:PEG2000318K8A10pc} 
\end{table}

\begin{figure}[h]
\centering
\includegraphics[width=180mm]{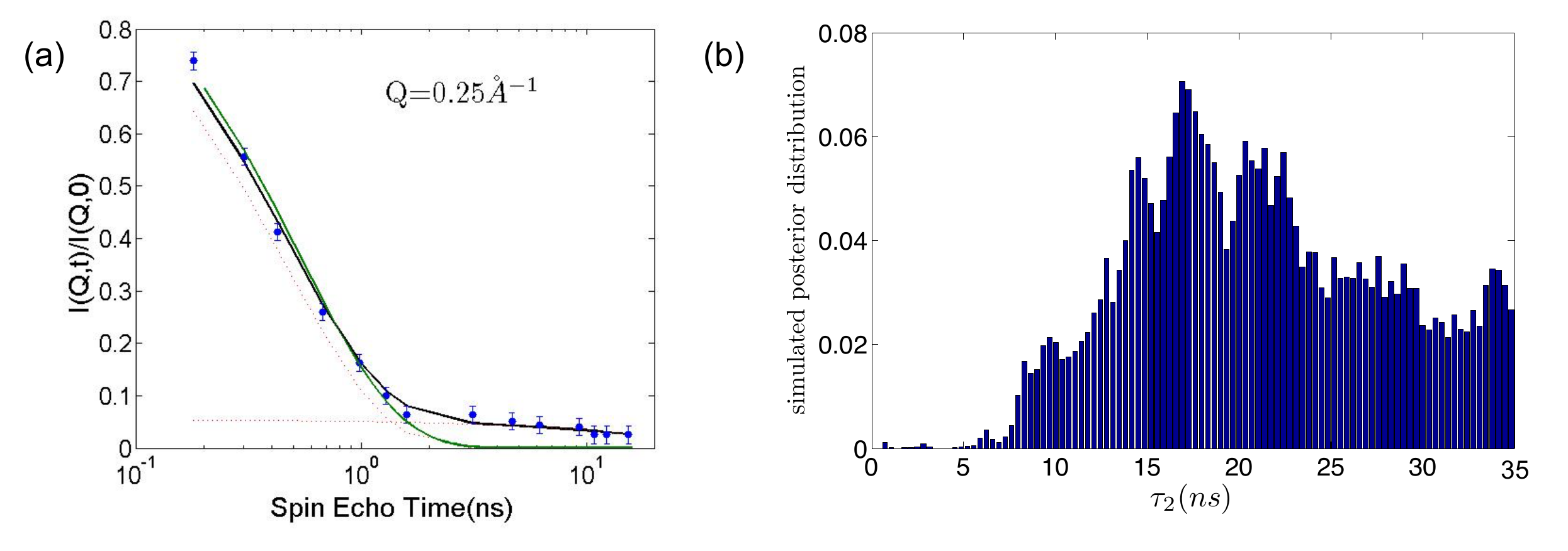}
 \caption{ (a) NSE curves representing  $I(Q,t)/I(Q,0)$ measured on the PEG2000 at 318 K,  $\lambda$ = 8 \AA, $Q$ = 0.25 \AA$^{-1}$ (symbols). The lines represent the best fits for the models with one (green continuous line) and two (dash black line) components; the dotted lines are the two contributions plotted separately corresponding to the model with two components. (b) Posterior distribution of the $\tau_2$ obtained by the Bayesian analysis.}\label{fig:PEG2000duefit}
\end{figure}

\clearpage

\section{PEG400}

\begin{figure}[htbp]
	\centering
	\includegraphics[width=180mm]{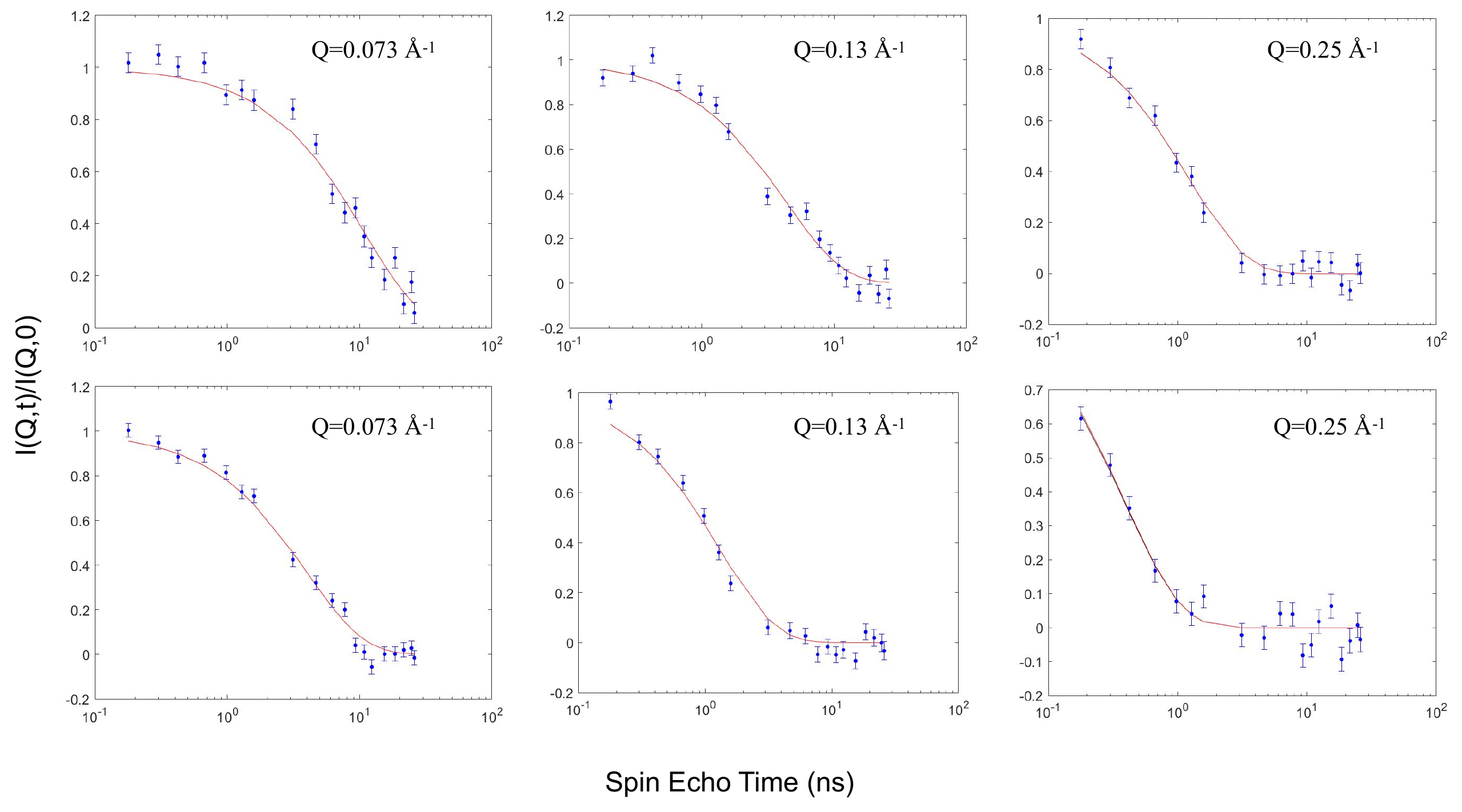}
	\caption{NSE curves representing $I(Q,t)/I(Q,0)$ measured on the PEG400 homopolymer 5\% concentration, at 280 K (top panels) and 318 K (bottom panels) at three selected $Q$ values and $\lambda$=8 \AA. The best fit curves (red continuous line) are shown.}	\label{fig:fitresultsPEG400}
\end{figure}

\begin{figure}[htbp]
	\centering
	\subfigure[]{\includegraphics[width=0.4\linewidth]{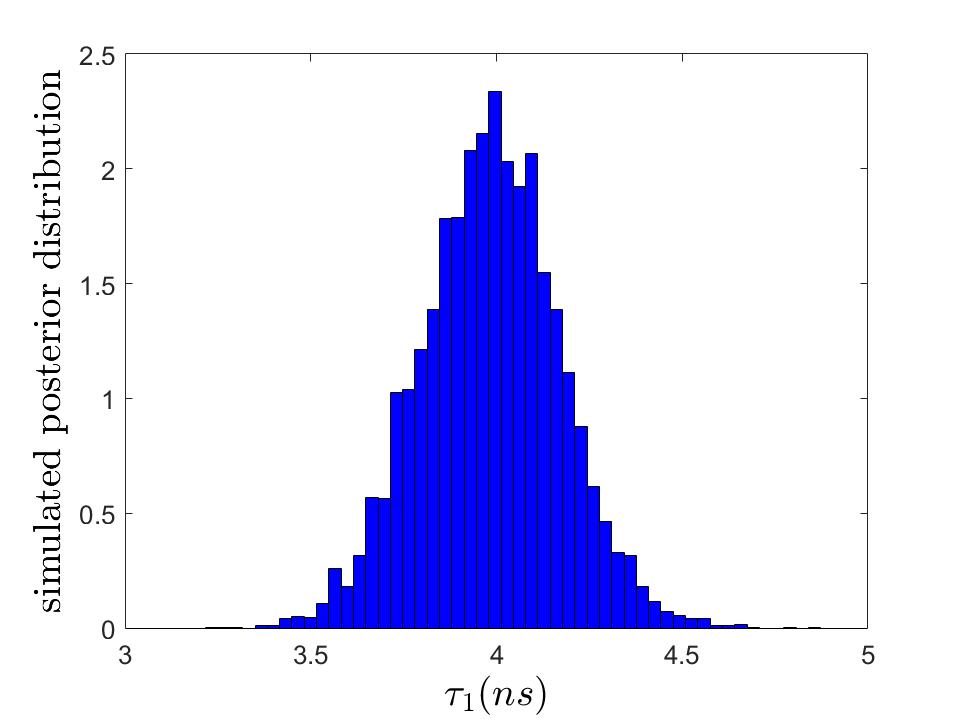}}	
	\subfigure[]{\includegraphics[width=0.4\linewidth]{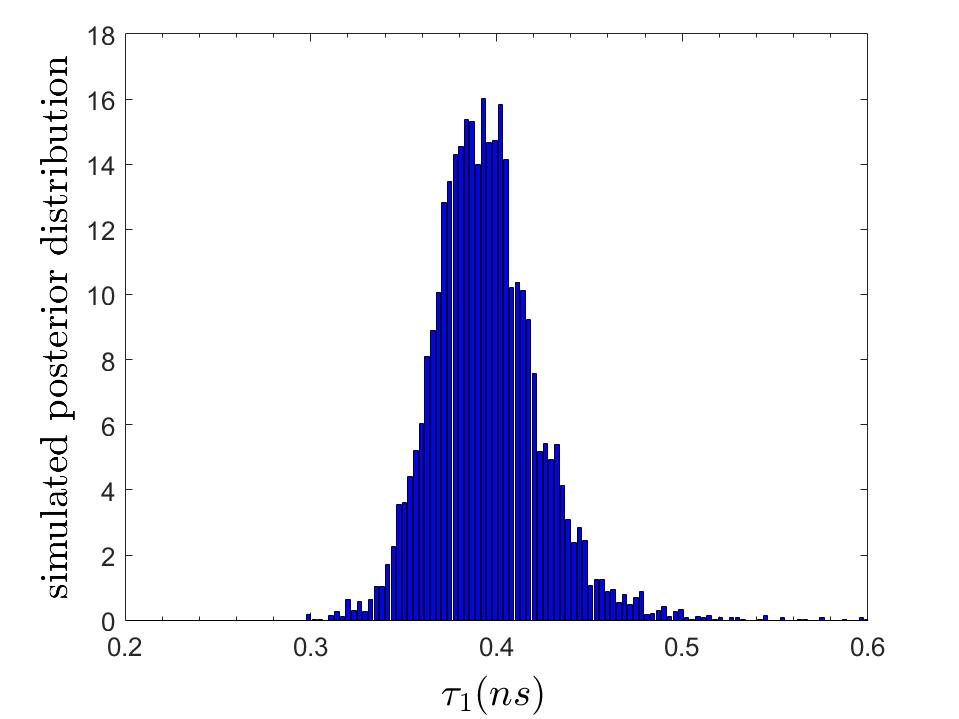}}		
		\caption{Two examples of the posterior distribution function of $\tau_{1}$ for the PEG400 homopolymers at 318 $K$, $\lambda$=8 \AA \ and $Q=0.073$ \AA$^{-1}$ (left) and $Q=0.25$ \AA$^{-1}$ (right).}\label{fig:fitresultspostPEG400}
\end{figure}

\clearpage
\section{Complementary information to the Bayesian Analysis}
\subsection{*The RJ-MCMC algorithm}
 The RJ-MCMC sampler we propose performs $M$ sweeps and, at each sweep $m$, all the parameters, including $k$, are updated in turn.
 This is performed by drawing the new values of a certain parameter conditionally on the data and all the other parameters.
 The algorithm uses different fixed-dimension moves to update the model parameters, conditionally on a fixed $k$, plus a variable dimension move to update the number of components $k$.

 In the fixed dimension moves, all the parameters, but $\nu$, are updated by means of Metropolis-Hastings moves \cite{Tierney94}. For a generic parameter $\theta$ to be updated, the Metropolis-Hastings move proposes, at each sweep $m$, a new candidate value $\theta^\prime$, drawn from a proposal distribution $q(\b\cdot|\theta)$, which depends on the current value of $\theta$ as well as from tuning parameters. This candidate value is then accepted with a probability calculated according to an acceptance rule, which ensures that the Markov chain converges to an equilibrium distribution which is the joint posterior of the parameters, and is given by $\min\{1,R\}$, where
 \begin{equation}\label{accprob}R=\frac{P(y|\Theta^\prime)}{P(y|\Theta)}\frac{P(\Theta^\prime)}{P(\Theta)}\frac{q(\theta|\theta^\prime)}{q(\theta^\prime|\theta)},\end{equation}
 with $\Theta^\prime$ denoting the whole parameter vector, in which the  parameter $\theta$ has been substituted by $\theta^\prime$, $P(y|\Theta)$ being the likelihood and $P(\Theta)$ the prior on the  parameters. Notice that the product of the first two ratios is simply the ratio between the posteriors evaluated in $\Theta^{\prime}$ and $\Theta$, respectively, with the normalizing constants cancelling out. The tuning parameters are tuned during some pilot running of the chain in such a way to guarantee an acceptance rate of the Metropolis-Hastings move of approximately $30\%$. To give an example, we describe in detail the updating of the component weights $A$, the Metropolis-Hastings updating of the other parameters being similar. At the $m$-th sweep of the algorithm, in order to update the current parameter $A$, we draw a candidate $A^\prime$ from a proposal $q(A^\prime|A)$. A reasonable and natural choice of the proposal distribution for $A^\prime$ is the Dirichlet distribution with vector of parameters proportional to the current value of $A$, i.e. $\mathcal{D}(\epsilon_AA_{1},\ldots,\epsilon_AA_{k})$, where $\epsilon_A$ is a tuning parameter. Under this choice, the proposed $A^\prime$ fulfils the constraints $A_j^\prime\ge 0, \forall j$, and $\sum_{j=1}^k A_j^\prime =1$, and the distribution of each $A_j^\prime$ is centred on the current $A_j$, with variance controlled by $\epsilon_A$. After drawing $A^\prime$, the acceptance probability is calculated as $\min\{1,R\}$, where, accordingly to Eq.(\ref{accprob}),
 $$R=\frac{\prod_{i=1}^n\phi\left(y_i;\gamma\sum_{j=1}^{k} A_{j}^\prime\exp\left(-\left(\frac{t_i}{\tau_{j}}\right)^{\beta_j}\right),\nu\sigma_i^2\right)}{\prod_{i=1}^n\phi\left(y_i;\gamma\sum_{j=1}^{k} A_{j}\exp\left(-\left(\frac{t_i}{\tau_{j}}\right)^{\beta_j}\right),\nu\sigma_i^2\right)} \times
 \frac{d(A^\prime;\lambda)}{d(A;\lambda)}
 \times\frac{d(A;\epsilon_A A^\prime)}{d(A^\prime; \epsilon_A A)},$$
 with the three ratios being, respectively, the likelihood, the prior and the proposal ratios in Eq.(\ref{accprob}), and $d(\b\cdot;\lambda)$ representing the density of the $\mathcal{D}(\lambda)$ distribution evaluated in a specific point ``$\b\cdot$''. Finally, to decide whether the move is accepted or not, a random number $u$ is drawn from a Uniform distribution on the interval $[0;1]$  and compared with $\min\{1,R\}$. If $u<\min\{1,R\}$, the move is accepted and $A^\prime$ becomes the new parameter vector, otherwise the move is rejected and the parameter vector stays unchanged.
 
 The parameter $\nu$ is instead updated by means of Gibbs sampling. The prior of this parameter is, in fact, conjugated with the likelihood and this results in a closed form for its full conditional distribution (i.e. the posterior distribution of the parameter, given the data and the value of all the other parameters). Thus, a new value of the parameter $\nu$ can be directly draw from its full conditional distribution, i.e.  \begin{equation}\label{Gibbs}\nu^{-1}\sim\mathcal{G}\left(\iota+\frac{n}{2},\varsigma+\frac{1}{2}\sum_{i=1}^n \frac{\left(y_i-\gamma\sum_{j=1}^{k} A_{j}\exp\left(-\left(\frac{t_i}{\tau_{j}}\right)^{\beta_j}\right)\right)^2}{\sigma_i^2}\right).\end{equation}
 
 Finally, updating the value of $k$ implies a change of dimensionality for the vectors of weights $A$, stretching parameters $\beta$ and relaxation times $\tau$. We accomplish this by introducing a RJ move, consisting in a random choice between the birth of a new component or the death of an existing one.
 The probabilities of the birth/death alternatives are $b_k$ and $d_k = 1 - b_k$, respectively, and depend only on the current value of $k$. Of course, $d_1 = 0$ and $b_{k_{\max}} = 0$; otherwise we choose $b_k = d_k = 0.5$, for $k = 2,\ldots, k_{\max}- 1$. The move starts with a random choice between birth and
 death, using probabilities $b_{k}$ and $d_k$. For a
 birth, we let $k^\prime=k^{m}+1$ and we pick a position $j^\prime$ at random among $1,\ldots,k^\prime$, with
 probability $1/k^\prime$ for the place to be occupied by the new
 component $j^\prime$. Then, a weight for the new component is drawn as $A^\prime_{j^\prime}\sim \mathcal{B}(1,k)$ and a new vector $A^\prime$ is obtained by rescaling the existing weights,
 using $A_j^\prime = A_j(1 - A^\prime_{j^\prime})$. Also, values of $\tau^\prime_{j^\prime}$ and $\beta^\prime_{j^\prime}$ are drawn from their respective priors
 $$\tau^\prime_{j^\prime} \sim \mathcal{U}(0,\tau_{\max})\qquad \mbox{and}\qquad \beta^\prime_{j^\prime}\sim \zeta\mathcal{B}(\kappa,\psi)+(1-\zeta)\delta_{\beta^\prime_{j^\prime},1}$$ and collocated in the $j^\prime-$th place of vectors $\tau$ and $\beta$ to give $\tau^\prime$ and $\beta^\prime$, respectively.
 The parameters $\nu$ and $\gamma$ remain unchanged.
 The acceptance probability for the birth move is calculated as $\min(1, B)$, where, accordingly to the RJ rule
 \begin{eqnarray}\label{birthdeath}
 B &=& \frac{\prod_{i=1}^n\phi\left(y_i;f^\prime(t_i),\nu\sigma^2_i\right)}{\prod_{i=1}^n\phi\left(y_i;f(t_i),\nu\sigma^2_i\right)}\times\frac{p(k^\prime)}{p(k)}\times \frac{d(A^\prime;\lambda)}{d(A;\lambda)} 
 \times  \frac{d_{k^\prime}}{b_{k}be(A^\prime_{j^\prime};1,k)}\times (1-A^\prime_{j^\prime})^{k}
 \end{eqnarray}
 where the first three factors in the multiplication represent the likelihood ratio (with $f(t_i)$ and $f^\prime(t_i)$ being the model function evaluated at the current parameter set and the new parameter set, respectively)  and the prior ratios, the fourth factor is the proposal ratio (with $be(\b\cdot;1,k)$ denoting the beta density evaluated in ``$\b\cdot$'') and the fifth factor is the Jacobian of the transformation from $A$ to $A^\prime$. Thus, the acceptance probability for the RJ move is obtained from the Metropolis-Hastings acceptance probability in Eq.(\ref{accprob}) simply multiplying by the Jacobian of the transformation from the old to the new parameters.
 For a death move, a component is chosen at random among any existing one, the chosen component is deleted and the remaining
 weights are rescaled to sum to 1. 
 The acceptance probabilities for a death move is $D=\min(1, B^{-1})$, with $B$ given in Eq.(\ref{birthdeath}).

\end{document}